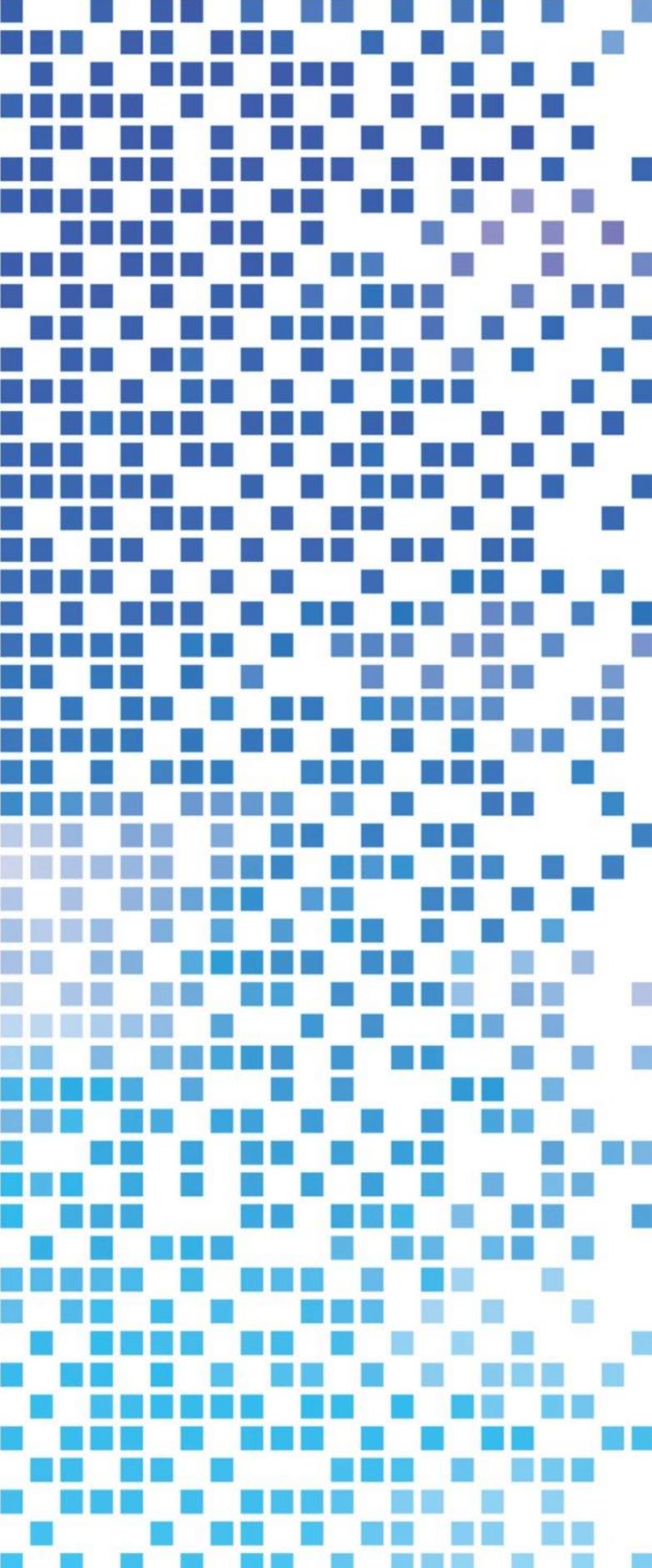

# ESCAPE

**Preparing Forecasting Systems for the Next generation of Supercomputers**

# D3.3 Performance report and optimized implementation of Weather & Climate Dwarfs on GPU, MIC and Optalysys Optical Processor

Dissemination Level: Public

This project has received funding from the European Union's Horizon 2020 research and innovation programme under grant agreement No 67162

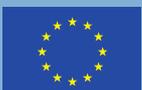

Funded by the
European Union

Co-ordinated by

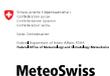 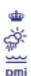 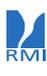 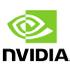 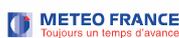 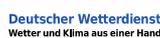 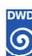 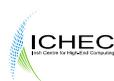 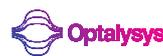 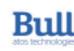 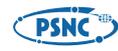 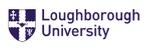

# ESCAPE

**Energy-efficient Scalable Algorithms**

**for Weather Prediction at Exascale**

Authors **Cyril Mazauric, Erwan Raffin, Xavier Vigouroux, David Guibert (Bull), Alex Macfaden (Optalysys), Jacob Poulsen, Per Berg (DMI), Alan Gray, Peter Messmer (NVIDIA)**

Date **2017/12/22**

## Table of Contents

## Figures

## Tables

# 1   Executive Summary

The goal of the ESCAPE project is to analyse and optimize time- and energy-to-solution for core components of numerical weather and climate simulation codes on modern hardware platforms. These components, so called Dwarfs, are extracted from existing weather and climate codes, ported to accelerated systems by domain scientists and optimized by the hardware experts so that they can ultimately be used as building blocks or design guidance for next generation numerical weather and climate code.

Here we summarize the work performed on optimizations of the dwarfs on CPUs, Xeon Phi, GPUs and on the Optalysys optical processor. We limit ourselves to a subset of the dwarf configurations and to problem sizes small enough to execute on a single node. Also, we use time-to-solution as the main performance metric.

Multi-node optimizations of the dwarfs and energy-specific optimizations are beyond the scope of this report and will be described in Deliverable D3.4.

To cover the important algorithmic motifs we picked dwarfs related to the dynamical core as well as column physics. Specifically, we focused on the formulation relevant to spectral codes like ECMWF's IFS code.

The main findings of this report are:

-   Acceleration of 1.1x - 2.5x of the dwarfs on CPU based systems using compiler directives
-   Order of magnitude acceleration of the dwarfs on GPUs (23x for spectral transform, 9x for MPDATA) using data locality optimizations
-   Demonstrated feasibility of a spectral transform in a purely optical fashion

In addition to these quantifiable results, the work performed here also lead to some best practices (e.g. investigate data transforms in deep call trees) that can be used for future optimization efforts.

While a direct comparison of CPU and GPU accelerated implementations for some dwarfs is feasible, most dwarfs will require multi-node optimizations to make any realistic comparisons. We therefore defer these comparisons mostly to Deliverable D3.4.

## 2   Introduction

### 2.1   Background

ESCAPE stands for Energy-efficient Scalable Algorithms for Weather Prediction at Exascale. The project develops world-class, extreme-scale computing capabilities for European operational numerical weather prediction and future climate models. ESCAPE addresses the ETP4HPC Strategic Research Agenda 'Energy and resiliency' priority topic, promoting a holistic understanding of energy-efficiency for extreme-scale applications using heterogeneous architectures, accelerators and special compute units by:

- Defining and encapsulating the fundamental algorithmic building blocks underlying weather and climate computing;
- Combining cutting-edge research on algorithm development for use in extreme-scale, high-performance computing applications, minimizing time- and cost-to-solution;
- Synthesizing the complementary skills of leading weather forecasting consortia, university research, high-performance computing centers, and innovative hardware companies.

ESCAPE is funded by the European Commission's Horizon 2020 funding framework under the Future and Emerging Technologies - High-Performance Computing call for research and innovation actions issued in 2014.

The objectives of ESCAPE are met by a cyclic collaboration of individual work package teams (WP): WP1 focuses on the extraction and definition of the dwarfs, WP2 on porting to accelerated architectures, WP3 on the optimization for individual target platforms. The insights and results gained from this work are then fed back into WP1 for a refined definition of the dwarfs.

In this report we present the work performed by WP3 on optimization of the dwarfs on alternative technologies and specialized accelerators such as NVIDIA GPUs, Intel Xeon Phi and Optalysys Optical Processors.

This work directly contributes to achieving the project's top-level objective 2, diagnose and classify weather & climate dwarfs on different HPC architectures, and 3, combine frontier research on algorithm development and extreme-scale, high-performance computing applications with novel hardware technology.

### 2.2   Scope of this deliverable

#### 2.2.1   Objectives of this deliverable

The goal of Deliverable D3.3 is to determine the optimal implementation and peak achievable performance of a subset of the dwarfs on select hardware. The dwarfs were selected to cover all major algorithmic motifs in a real-world application, including global transforms and column based models. In this first step, only single node performance was investigated. Also, energy efficiency was addressed by minimizing the overall runtime of the dwarf on a given hardware architecture.

Multi-node optimizations and energy related optimizations are deferred to Deliverable D3.4.

The deliverable consists of this report and software containing the platform specific optimizations for a selection of weather and climate dwarfs. This software is accessible via the ESCAPE Software Stash.

### 2.2.2 Work performed in this deliverable

The work performed for this deliverable consisted mostly of task 3.3 of the ESCAPE Description of Action (DoA). This includes optimizations of the dwarfs adapted to CPUs, Xeon Phi, GPUs and the Optalysys optical Processor.

Starting from the portable implementations developed in WP2, the team members of WP3 took these dwarfs and optimized them for their respective platform. The approaches differed, based on platform due to the maturity of the implementation of the dwarfs for the respective platforms. For example, the CPU implementation was already widely used in production and expected to execute relatively efficiently, whereas the implementation on GPUs were relatively new and significant optimization potential was expected. Finally, the implementation on the optical processor was more at the level of a proof of concept study, rather than a complete implementation of the dwarf.

One of the challenges of optimization work is to determine the point of diminishing return. Using theoretical models like the roofline model to assess the current performance related to the hardware limits can therefore help to determine the performance ceiling. Nevertheless, even with theoretical guidance it is often a lengthy process to discover and implement further optimizations. Consequently, the results presented here need be considered a snapshot in time of this ongoing optimization work and we expect future optimizations, e.g. related to multi-node optimizations, to impact these results as well.

In addition to arriving at an optimized implementation of specific dwarfs, the project aims at determining best practices for implementing numerical methods on different architectures.

### 2.2.3 Deviations and counter measures

There were no deviations from the original proposal.

### 2.2.4 Organization of this report

In the following chapters we summarize the main results of the optimization effort. We will then present the work performed on the individual dwarfs, including Spectral transform, MPDATA, Radiation, Bi-Fourier, Cloud Microphysics and Elliptic solver GCR. Each chapter gives some background on the dwarf and then describes the work performed and results obtained for each platform. Specifically, we focused on the Spectral transform dwarf as it is relevant to spectral codes like ECMWF's IFS one.

## 3   Result highlights

Here we summarize the main findings of these investigations. Due to the different nature of the architectures, different approaches to optimization were used: For the CPU based architectures, we used low-intrusion optimizations based on directives, for GPUs we optimized the code via restructuring and finally the Optalysys optical processor required more fundamental considerations.

The CPU version of the code is currently used operationally and the assumption is that the code is already in an advanced optimization state. The optimization work therefore concentrated on the use of directives to support the compiler generating better vectorized code and therefore better utilizing modern CPU architectures. In addition, different methodologies for using the cores on a CPU socket were investigated, and it turned out that the low-overhead OpenMP offered significantly better performance than an approach relying on explicit message exchanges via MPI. Overall, these optimizations resulted in a speedup of 1.1x - 2.5x for the dwarfs investigated, both for CPUs and Xeon Phi co-processors.

For the GPU, much more in-depth optimizations were performed. This included a detailed analysis of the data motion and data representations, and eliminating unnecessary transpositions or data reformatting. As a result, the dwarf was accelerated by more than a factor of 23x, utilising more than 40% peak performance of the GPU according to the roofline model. With a similar approach, the MPDATA dwarf was accelerated by a factor of 9x.

These optimizations are labour intensive, but demonstrate the performance potential for the algorithms on modern architectures. It is expected that these optimizations will also benefit a CPU based implementation.

Finally, the feasibility to execute the core algorithms on an optical processor were investigated. As it turns out, the implementation is indeed feasible, but questions about the data representation, as well as speed for the input and read-out electronics need to be improved to become competitive with existing silicon based computing hardware.

## 4   Dwarf 1: Spectral Transform - Spherical Harmonic

### 4.1   Introduction

In the following sections we will describe the optimization strategies and performance results of the spectral transform dwarf on the different hardware platforms.

A detailed description of the physics of this dwarf is presented in the dwarf documentation, available on the ESCAPE website.

### 4.2   GPU Optimizations

In this section we present our optimization work of the spherical harmonics spectral transform ESCAPE dwarf (which uses a spherical grid for global simulations) on the NVIDIA GPU architecture, and corresponding performance results. For this work, we use the "prototype 2" version of the dwarf, and the relatively small TL159 test case (which features 125km resolution), since it fits on a single GPU. The larger test cases require multiple GPUs, and multi-GPU aspects will be presented in a later deliverable. We use the NVIDIA Tesla P100 GPU for all results. As will be seen, the TL159 case is large enough such that the kernels can typically achieve a good percentage of peak performance on the GPU. Therefore, the work and results presented below, where execution of the whole grid takes place on a single GPU, are largely transferable to the computational aspects of execution of larger cases on multiple GPUs (with a sub-grid executing on each GPU). We first describe the steps taken to optimize the code, and go on to present performance results in the context of the Roofline performance model for the GPU in use.

### 4.2.1 Roofline model

The line in Figure 1 is known as a "Roofline", and provides an upper limit for achievable application performance according to the Roofline model. This simple but effective model classifies codes in terms of their operational intensity (OI) (also known as arithmetic intensity), given by the ratio of floating point operations (in flops) to volume of memory accesses (in bytes). For relatively low OI, the performance (in flops) is limited by the available global memory bandwidth of the architecture, which has been measured using the STREAM benchmark and is represented on the plot by the sloping part of the line. For relatively high OI: the limiting factor is the peak compute capability of the hardware: this is given by the flat part of the line. The "kink" in the line is at the specific OI where the regimes cross over: this is known as the "ridge point" and is given by ratio of peak flops to available memory bandwidth of the hardware. In other words, the ridge point it is a measure of how many operations can be performed per byte loaded when the chip is running at peak performance.

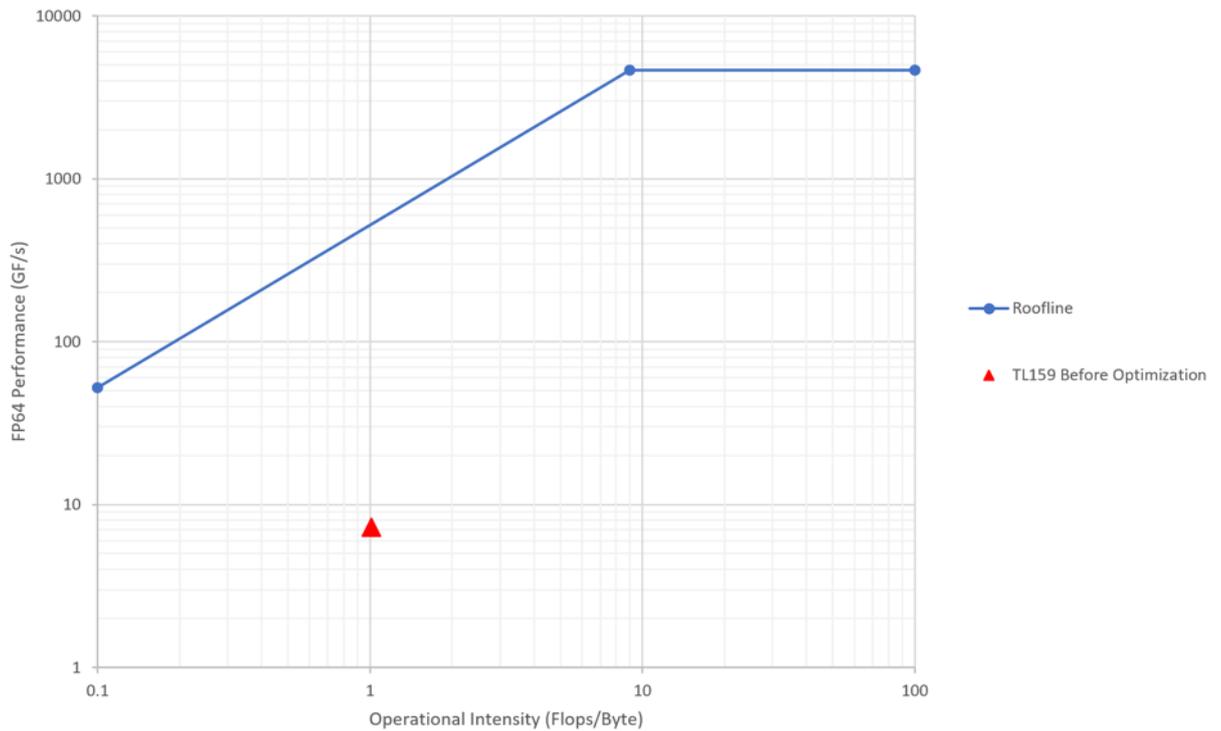

*Figure 1: The performance of the original version of SH dwarf TL159 test case on the NVIDIA Tesla P100 GPU, in the context of the roofline for this architecture. The point is positioned in the plot according to its operational intensity: points under the sloping region of the roofline are limited by available memory bandwidth, and points under the horizontal region are limited by peak computational performance.*

### 4.2.2 Baseline Performance

Before the work presented in this report, there existed a GPU-enabled version of the code using OpenACC and CUDA libraries, and the performance of this is included in Figure 1. It can be seen that the original code is performing between 1 and 2 orders of magnitude lower than the roofline, indicating the clear scope for optimization. Before describing the optimizations, we will show why the performance of the original case is so poor.

A portion of the timeline of the original code, as given by the NVIDIA profiler, is shown Figure 2. From the window on the left, it can be seen that the profile contains a huge number of small kernels. Note that the timeline shown just corresponds to a portion of the full simulation time-step, but the pattern is similar throughout. We can get some more clarity by zooming in further, as shown in the right window. It can be seen that the time spent performing actual computation (as represented by the fraction of the "Compute" row containing solid blocks) is very low. Instead, most time is spent in "Driver API" calls. These are overheads mainly associated with memory allocation, data movement and CUDA kernel launches. To improve performance, it is imperative to restructure the code to contain a much lower number of larger kernels, and to remove these overheads.

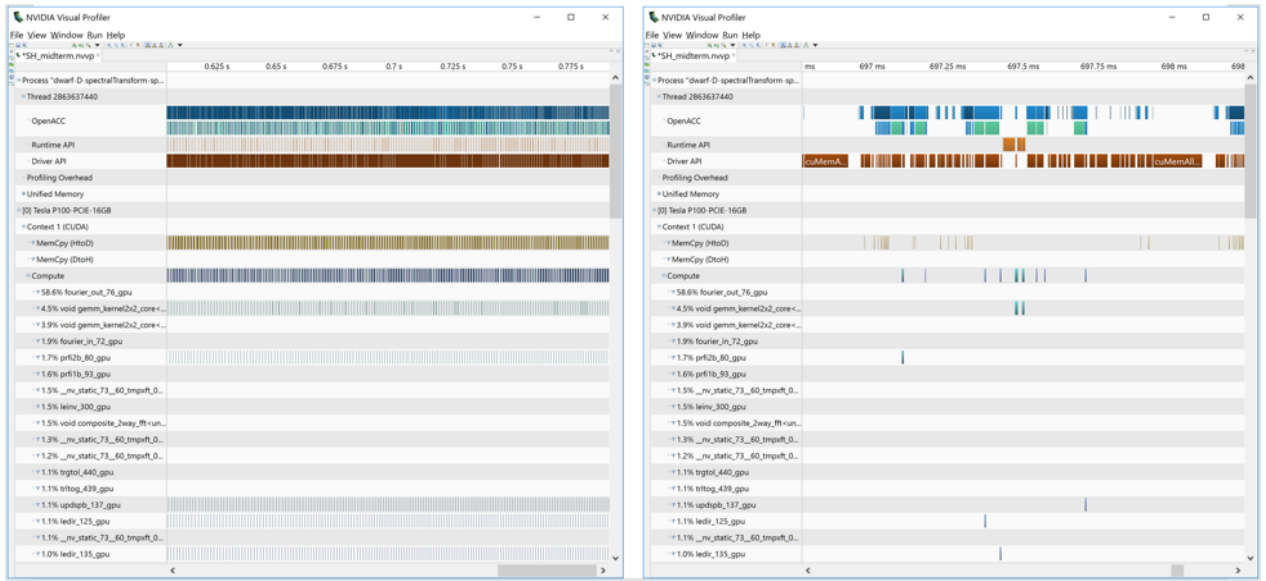

*Figure 2: The timeline of the original code for the TL159 test case, as given by the NVIDIA profiler. Left: shown is a portion of the simulation time-step, which can be seen to contain a very large number of small kernels. Right: a zoomed portion of figure on the left, where it can be seen that the time spent in "Compute" is very sparse, and instead the timeline is dominated by API calls.*

The operations in the original code, in essence, have the following structure (albeit spread throughout multiple files and subroutines):

```
Loop over time-steps
    …
  Loop over first spherical dimension
       OpenACC memory allocs, frees, copies, etc
             (scattered throughout this loop body)
       OpenACC loops over 2nd dim (specific to outer loop) and fields
             operation
       …
       CUDA Library calls
       …
       more OpenACC loops similar to above
       …
```

The simulation notionally has three levels of parallelism: two from the spherical dimensions of the grid (latitude and longitude), and a third from use of multiple "fields" associated with multiple altitude levels. It can be seen that only two of these were being exposed to the GPU, with a sequential loop corresponding to the other spherical dimension at the outermost level. Furthermore, within this outermost loop, there are numerous expensive data management operations scattered throughout.

### 4.2.3 Optimization Strategy

The main optimization strategy involves restructuring the above as follows:

```
OpenACC memory allocs (reusable arrays), copies

Loop over time-steps

    …

    OpenACC loops over 1st dim, 2nd dim and fields

        operation

        …

    CUDA Library calls (with batching where possible)

        …

    more OpenACC loops similar to above

    …

OpenACC frees
```

It can be seen that all levels of parallelism are now exposed to the GPU, and that the expensive data management operations have now been moved to the outermost level so don't occur every time-step, with the introduction of reusable arrays. This allows the computational operations to be completed by a much lower number of larger kernels than previously possible, with kernel launch overhead being vastly reduced.

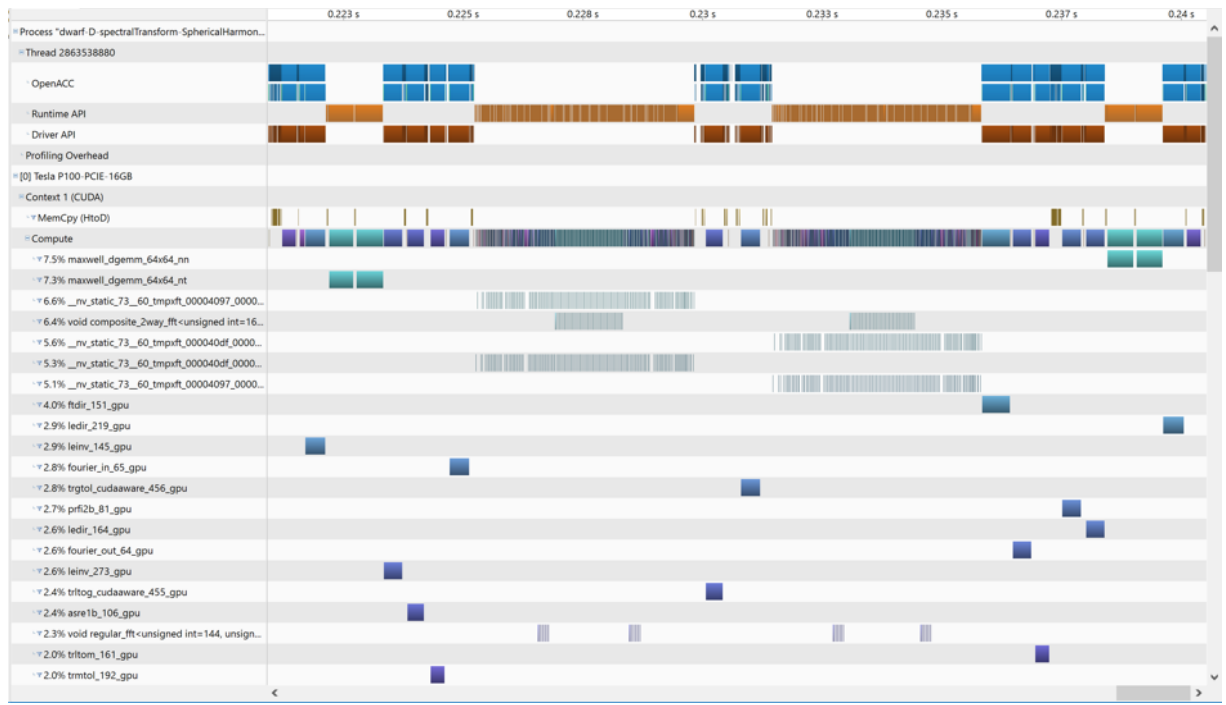

*Figure 3: The timeline of the optimized code for the TL159 test case, as given by the NVIDIA profiler. The full simulation time-step is shown. It can be seen that there now exists a much lower number of larger kernels, and the "Compute" line is now much more highly packed.*

In Figure 3 we show the timeline of the optimized code, for comparison with Figure 2. The new timeline shows a full simulation time-step. It can clearly be seen that we now have a much smaller number of larger kernels; the "Compute" row is now almost completely packed with blocks; and the timeline is no longer dominated by API overhead. The time taken by this optimised version is 23x lower than the original version (as will be discussed in more detail later).

We now discuss in more detail the work done to adapt the different types of kernels: matrix multiplications and Fourier transforms (both performed using CUDA library calls), and OpenACC kernels.

### 4.2.4   Matrix Multiplications

Matrix multiplications are required within both the direct and inverse Legendre Transforms. Originally, there existed many library calls to the cuBlasDgemm library, each responsible for a relatively small matrix multiplication: this suffered from low parallelism exposure and launch latency sensitivity. Our restructuring provides the opportunity to expose the parallelism of the other spherical dimension to the library. However, there is a complication that (due to the nature of the spherical grid) the size of each multiplication varies across this dimension. There exists a "batched" version of the library, but unfortunately it does not support differing sizes. But due to the nature of matrix multiplication, it is possible to perform each operation using arbitrary large matrices, provided that the "extra" entries contain zero values. That is, we can pad each matrix with zeroes up to the largest size, and operate using the standard batched library call with a uniform batch of matrices of that size, where each will be padded (except the largest).  Extra computations occur, but have no effect on the result since they do not contribute to the dot-product accumulations for each element of the result matrix, despite these extra computations, the overall performance is much improved.

### 4.2.5   Fourier Transforms

In a similar fashion to the matrix multiplications described above, calls to the CUDA FFT library are required within both the direct and inverse Fourier Transform stages of the time-step. These operate in a single spherical dimension, and can be batched over fields in a relatively straightforward manner using the batching interface to the library. Again, this results in many small calls with low parallelism exposure and launch latency sensitivity. The code restructuring facilitates exposure of parallelism associated with the other spherical dimension.  However, the varying sizes across this spherical dimension are, again, not supported by the batched library. They constitute more of a problem for FFTs, because we cannot pad with zeros in the same way we did with matrix multiplication (which would lead to different results). However, we have worked around launch latency problem by removing CUDA device synchronization after each call, such that the CUDA driver is able to overlap launch latency with execution. In the future, further improvements may be possible through development of a custom FFT that supports batches with differing sizes, such that the parallelism can be fully exposed on the GPU.

### 4.2.6   OpenACC kernels

#### 4.2.6.1 Algorithmic Structure

The original code resulted in the execution of many small hand-coded OpenACC kernels, again with low parallelism exposure and with significant launch overheads. The code restructuring has allowed the other spherical dimension to be folded into each OpenACC kernel; this combination means we now have a much smaller number of larger kernels, each typically with the following structure:

```
Loop over 1st spherical direction
    …
    Loop over 2nd spherical direction
        …
        Loop over fields
            …
            Operation involving multidimensional arrays
```

#### 4.2.6.2 Parallel Loop Structure

There are different options as to how to parallelize this using OpenACC. The naïve option is to add a "parallel loop" construct to the outermost loop, and "loop" constructs to each of the innermost loops. However, this often results in suboptimal performance since only the innermost loop will be mapped to CUDA threads, and performance will only be optimal when the extent of this loop (determined by the physics) provides a good map to the thread block size. Where possible it is better to restructure the loops such that they are perfectly nested and use the "collapse(3)" clause such that the loops are collapsed into a single loop by the compiler, therefore the parallelism from all three loops is mapped to all levels of the hardware, which is much more flexible:

```
!$acc parallel loop collapse(3)
Loop over 1st spherical direction
    Loop over 2nd spherical direction
        Loop over fields
            …
```

A complication is that, due to the spherical grid, the extent of the second loop is dependent on the first loop, which prevents such restructuring/collapsing. However, this can be enabled through the following adaptation:

```
!$acc parallel loop collapse(3)
Loop over 1st spherical direction
    Loop over MAX EXTENT of 2nd spherical direction
        Loop over fields
            Work out real extent, and mask using an if statement
                …
```

This means that certain chunks of threads are never active for the main operation, but since these chunks are contiguous this typically does not have a major effect of performance (as will be seen).

### 4.2.6.3   Memory Access Patterns

All the OpenACC kernels have an OI in the memory-bandwidth bound regime, so access patterns are extremely important in achieving a high percentage of available memory bandwidth. With the above parallel loop structure, it is important that the memory access patterns into the multidimensional arrays result in memory coalescing: this is achieved when the innermost loop index corresponds to the first (innermost) dimension of the Fortran array. We choose memory layouts for our temporary scratch arrays that correspond to this behaviour. In some kernels, we originally had patterns corresponding to matrix transposes, such that either the reading or writing was in the wrong order for coalescing. For some, coalescing was facilitated through replacing $C = AB$ matrix multiplications by equivalent $C^T = B^T A^T$. This allows transpose operations (which are associated with poor memory access patterns in naïve OpenACC code) to be pushed into the DGEMM library calls, which have much higher-performing implementations of transposed data accesses. There remain transpose patterns within kernels involved in transposing grid point data from column structure to latitudinal (and inverse) operations, which naturally involve transposes and are thus harder to fix through restructuring. However, we optimized these using the "tile" OpenACC clause, which instructs the compiler to stage the operation through multiple relatively small tiles which can perform the transpose operations within fast on-chip memory spaces, such that the accesses to global memory are much more regular.

### 4.2.6.4   Replacement of Deep Data Accesses

Another optimization involved data structures that remain constant throughout the execution of each kernel with read-only accesses. When programming directly in CUDA, the normal strategy would be to load these into the fast on-chip constant cache before kernel launch, to avoid unnecessary global memory accesses. Ideally, when using OpenACC, the compiler would automatically perform this task, but it was found that this was not being done for the many variables that reside within "deep" data structures. For example, there exists a variable "D" which is a Fortran derived type containing information about the distributed memory environment. This contains multiple variables, e.g. NUMP (a scalar representing the number of spectral waves handled by the processor) and MYMS (an array containing the corresponding wave numbers). These can be accessed in kernels using the standard Fortran syntax, e.g. D%NUMP or D%MYMS[i], but the compiler does not utilize constant memory and the performance is suboptimal (e.g. the degradation we measured at 25% for one kernel). Instead, we use equivalent "flat" structures e.g. D_NUMP and D_MYMS, for which the compiler does use constant memory and we get automatic constant memory usage, with performance closer to the memory-bandwidth roofline. Of course, this requires extra code to allocate and free memory both on the CPU and GPU, copy data from the deep to flat structures, and transfer to the GPU, but this is executed only once before the time-steps commence.

### 4.2.7   Performance Results

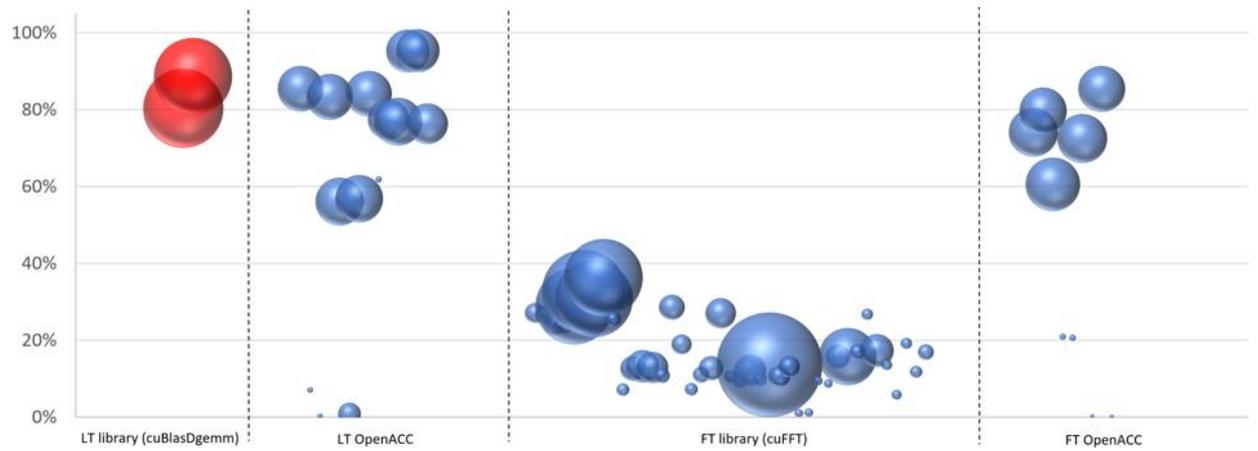

*Figure 4: The performance of each kernel in the TL159 case, given as a percentage of the NVIDIA P100 GPU roofline, where kernels have been classified according to the section of the time-step in which they reside. The size of each bubble is proportional to the time taken by that kernel. For red bubbles, the roofline is given by the peak computational performance of the GPU, for blue bubbles the roofline is achievable memory bandwidth as determined using the STREAM benchmark.*

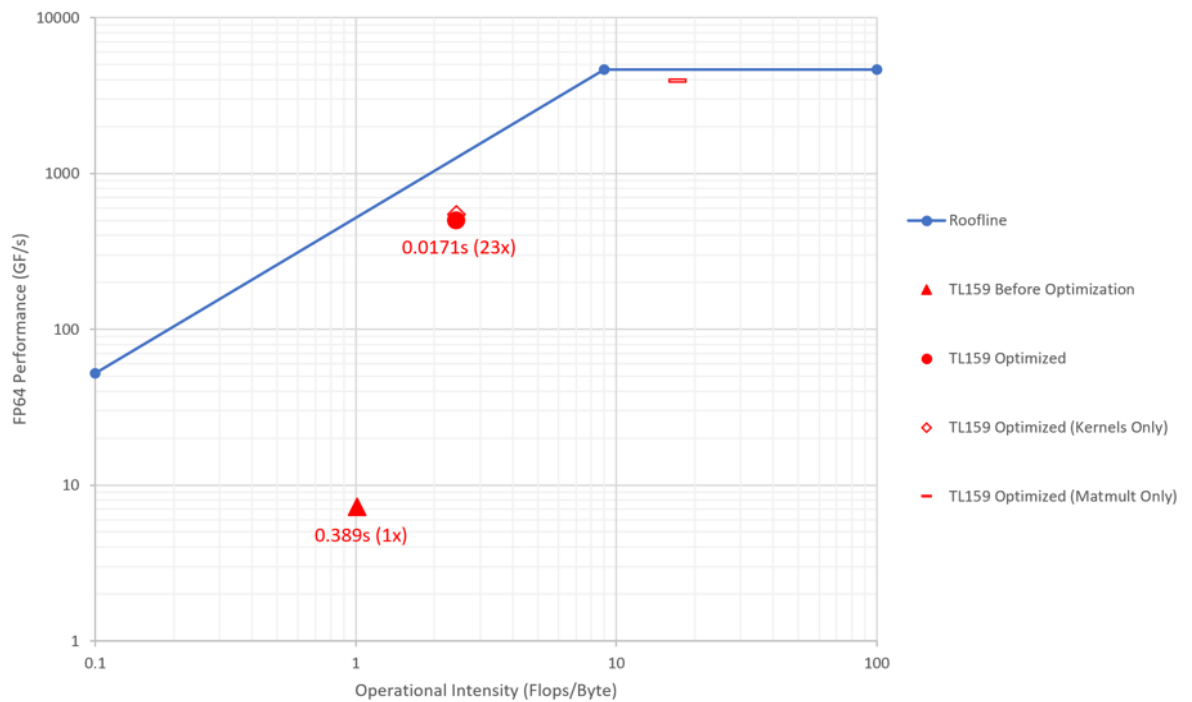

*Figure 5: The performance of the SH dwarf TL159 test case on the NVIDIA Tesla P100 GPU, in the context of the roofline for this architecture. The full time-step of the original code is represented by the solid triangle. The corresponding time-step for the optimized code is represented by the solid circle. Also included are partial results for kernels only (open diamond) and matrix multiplication only (open rectangle). Each point is positioned in the plot according to its operational intensity: points under the sloping region of the roofline are limited by available memory bandwidth, and points under the horizontal region are limited by peak computational performance.*

In Figure 4, we show the performance of each kernel individually where we have classed them into, from left to right: Legendre Transform matrix multiplication library calls, Legendre Transform OpenACC kernels, Fourier Transform library calls, and Fourier transform OpenACC kernels. The size of each bubble is proportional to the runtime of that kernel, and the height of the bubble corresponds to the percentage of roofline performance for that kernel (where the roofline is OI dependent and determined by the peak compute capability for red bubbles and achievable memory bandwidth as given by the STREAM benchmark for blue bubbles). Therefore, the higher the better and the larger the bubble the more important in terms of overall performance: small bubbles have limited significance and can be ignored. It can be seen that the matrix multiplications are performing very well at above 80% of peak performance, thanks to the batching described above. Similarly, the optimizations to the OpenACC kernels have resulted in most kernels being achieving 70-90% of achievable bandwidth. There are a few at around the 60% level, which is still reasonable, but indications from the profiler are that these are being limited by addressing operations associated with those parallel threads that do not end up performing any operations (see above). The main limiting factor for performance is seen to be the calls internal to the FFT library, which are seen to achieve around 10-40% of peak bandwidth. Development of custom FFT code which allows batching of different sizes, for use in place of the standard library, may allow further improvements and that will be considered in future work.

The overall performance results are added to the Roofline plot in Figure 5. The time taken by the optimized code is 23x lower than the original and is achieving around 40% of roofline performance. The graph also includes the result for "kernel only" timings where API calls have been removed: the fact that this is only very slightly higher than the full time-step result confirms that we have largely eliminated API overhead. The performance of the matrix multiplications in isolation is also shown in Figure 5. Note that all performance figures include the extra operations on the zero values which occur due to the padding of the matrix multiplications. It can be seen that the matrix multiplication performance is higher than the overall performance (in flops) and the OI is moved to the right, into the compute-bound regime. Note that matrix multiplication is associated with $O(N^3)$ computational complexity for $O(N^2)$ memory accesses, where due to the extra padding operations, N is larger than previously.

### 4.2.8 Summary

In this chapter, we described optimizations that enabled the time taken by the single-GPU TL159 test case of the spherical harmonics spectral transform dwarf to reduce by a factor of 23. The algorithmic restructuring, kernel and library batching, removal of launch overheads and other kernel-level optimizations have allowed performance to reach around 40% of achievable peak (as given by the roofline model). Most components are in the 70-90% range, the main limiting factor is now FFT operations which currently utilize the CUDA FFT library. Development of custom FFT code which allows batching with differing sizes may further improve performance. The work presented is relevant for achieving best possible performance on each GPU for larger test cases distributed across multi-GPU: for these it is also important to optimize communication performance, which will be a main focus in our work going forward.

### 4.3 CPU Optimizations

#### 4.3.1 Baseline performance

The initial performance of the spherical harmonics spectral transform ESCAPE dwarf has been strongly benchmarked (with and without the use of the Atlas library) on different Intel Xeon and Xeon Phi processors. Particular attention has been paid to the intra node scalability aspect of the dwarf using OpenMP and MPI parallelisation.

Concerning the test cases used, a focus on large ones has been chosen when dealing with Xeon processors and smaller ones for Xeon Phi as less memory was available on the system. Two interesting test cases, TCO639 and TCO1279 have been used to compare Xeon and Xeon Phi. The latter test case represents the high resolution operational model at ECMWF (about 16 km resolution).

##### 4.3.1.1 Intra node scalability study: Atlas vs non-Atlas

This early study has been conducted on the original version of the dwarf using the processor, with the main characteristic depicted in Table 1, on a dual socket configuration. It can be noticed that no HyperThreading has been used for this experiment and a simple binding: each OpenMP thread or MPI task bound on one physical core (thread/task 0 on processor 0, thread/task 1 on processor 1, etc.).

Figure 6 shows the gain offered by the Atlas version compared to the non-Atlas one. As one can see, the use of the Atlas library provides better performance and especially a better scalability, as the gain grows according to the number of threads used (i.e. number of cores). Thus the rest of the presented results have been obtained using the Atlas based prototype of the dwarf.

| Processor type (number) | E5-2650V3 |
|---|---|
| Number of cores | 10 |
| Number of threads | 20 |
| Base / Turbo frequency | 2.3 / 3 GHz |
| Cache | 25 MB SmartCache |
| Bus Speed | 9.6 GT/s QPI |
| Thermal Design Power (TDP) | 105 W |
| Memory installed | 64 GB DDR4 2133 MT/s |

*Table 1 - Main characteristics of the Intel processor used for early intra node scalability.*

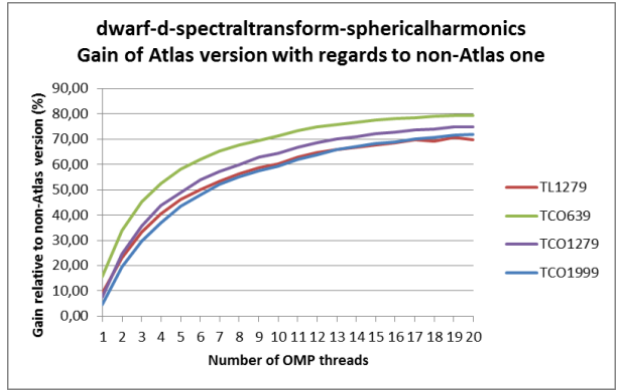

*Figure 6 - Performance gain of Atlas prototype versus non-Atlas one of the Spherical harmonics spectral transform dwarf according to the number of OpenMP threads (i.e. processors).*

### 4.3.1.2   Intra node scalability study: OpenMP vs MPI

The scalability results for the OpenMP and the MPI original versions of the dwarf are presented in Figure 7 and Figure 8 respectively. The OpenMP gives better performance than the MPI version, especially when the number of threads/tasks is high. The intra node scalability is better and more regular on OpenMP than MPI which figures out that the dwarf takes advantage to the shared memory execution model. But at large scale (i.e. many cores, multi nodes) we can expect better scalability for the MPI version which is its purpose.

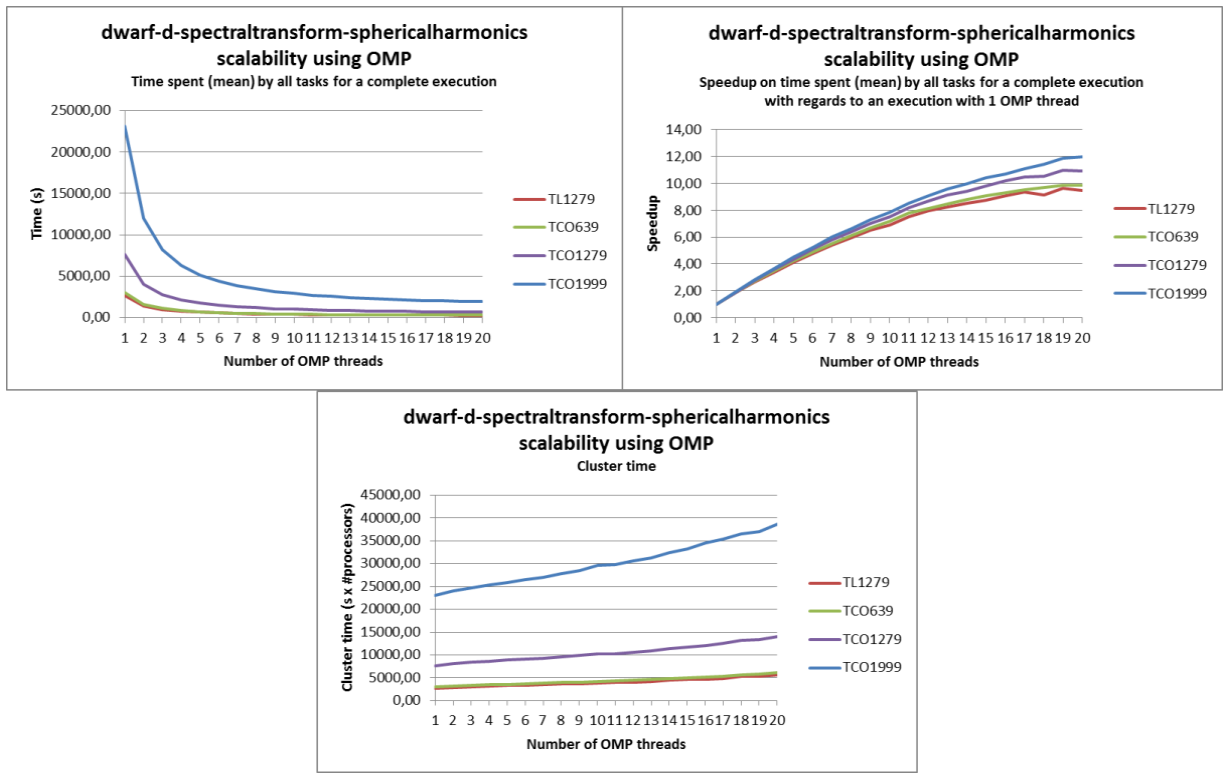

*Figure 7 - Spherical harmonics spectral transform dwarf intra-node scalability with OpenMP. Top left graph shows timing results, top right graph shows the speedup with regard to sequential execution and the bottom graph shows the cluster time (ideal cluster time is a constant, i.e. horizontal line).*

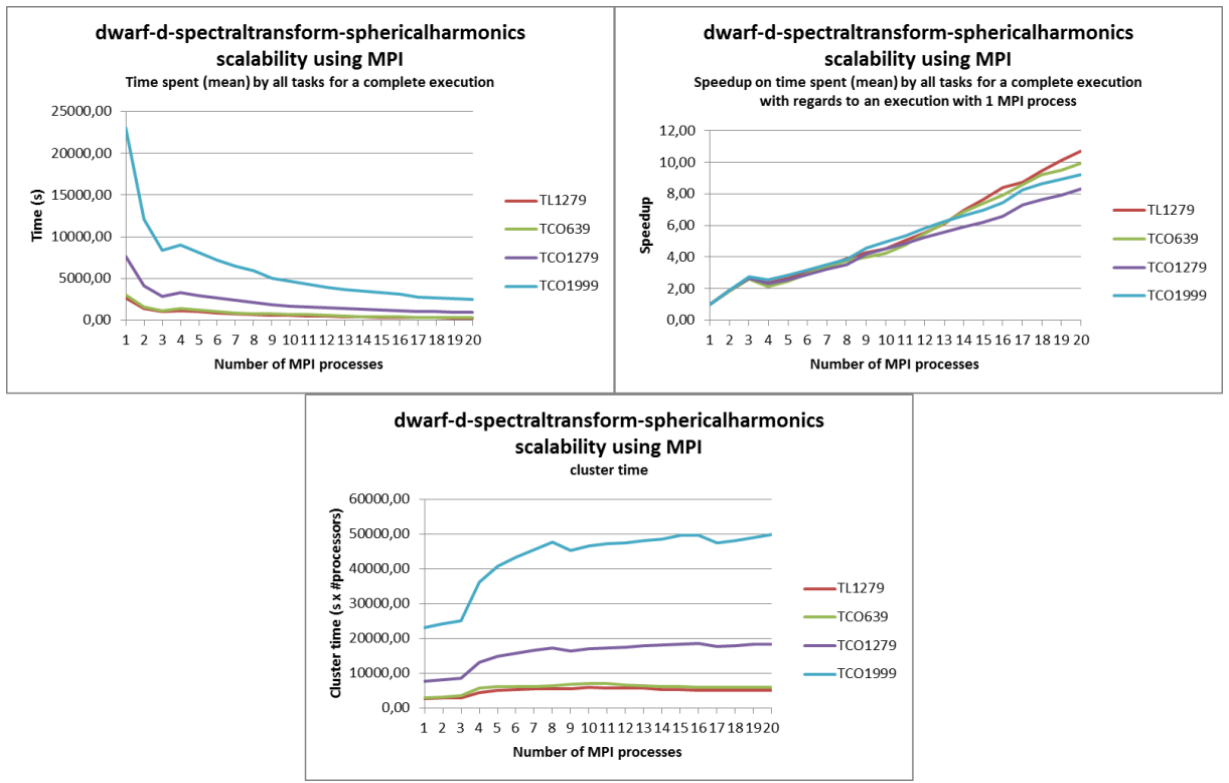

*Figure 8 - Spherical harmonics spectral transform dwarf intra-node scalability with MPI. Top left graph shows timing results, top right graph shows the speedup with regard to sequential execution and the bottom graph shows the cluster time (ideal cluster time is a constant, i.e. horizontal line).*

### 4.3.1.3 Profiling hybrid version configuration

The first analysis presented in previous section has been extended by a deeper profiling on a hybrid configuration (i.e. using both MPI and OpenMP). This work has been done on the same node as previously (see Table 1) using 10 MPI tasks and 2 OpenMP threads per MPI tasks (using all physical cores). Figure 9 shows that the OpenMP parallelization is high representing 70% of the execution time but also highlighted an average communication of 25% and a high MPI imbalance due to the use of "barrier" routine.

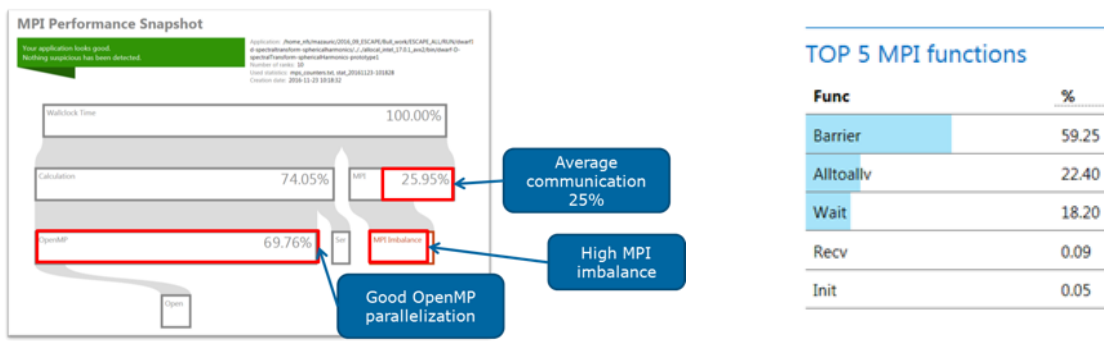

*Figure 9 - Profiling result of a hybrid configuration of the dwarf (from Intel MPI Performance Snapshot)*

### 4.3.2 Optimization strategy

According to profiling results, it clearly appeared that the main computational intensive kernels are the FFT and matrix multiplication executed by a dedicated highly tuned library (as Intel Mathematics Kernel Library, called MKL).

So before deeply restructuring the code, the first optimization strategy concentrated the effort on non-intrusive optimizations which have the advantage to be easier portable and maintainable. Among these optimizations, the use of extensions to the x86 instruction set architecture (ISA) as SSE, AVX, AVX2, AVX-512 is interesting as it gives the information on the capacity of the original dwarf source code to be vectorized. Then, when the compiler failed at vectorizing some loops or loop nests, more deep investigation has been done to guide the compiler using compiler directives. As the different instruction sets are not supported by all processors, the study proposed an intra node scalability comparison study among several available systems (at the time of benchmarking).

### 4.3.3 Performance results

In this part, a comparison of different configurations is presented. For this comparison several processors, instruction set, OpenMP/MPI and memory system configurations have been tested. The following table (Table 2) presents the main characteristics of the processors used.

It should be noticed that for the Intel Xeon Phi 7120p (noted KNC) which is an accelerator, an effort has been made as the dwarf was cross-compiled. This specificity required a different compilation strategy which was not well supported by the original compilation toolchain. Another remark: the dwarf has been executed natively (in contrast of an offload execution of the kernel where part of the code is executed on the host) which required a dedicated runtime.

| Processor type (number) | E5-2650 v3 | E5-2680 v3 | E5-2690 v4 | E7-8890 v4 | 7120p KNC | 7250 KNL |
|---|---|---|---|---|---|---|
| System type | Dual socket | Dual socket | Dual socket | SMP with 4 modules | Mono socket | Mono socket |
| Number of cores | 10 (x2) | 12 (x2) | 14 (x2) | 24 (x 4) | 61 | 68 |
| Number of threads | 20 (x2) | 24 (x2) | 28 (x2) | 48 (x 4) | 61 | 272 |
| Base / Turbo frequency | 2.3 / 3 GHz | 2.5 / 3.3 GHz | 2.6 / 3.5 GHz | 2.2 / 3.4 GHz | 1.24 / 1.33 GHz | 1.4 / 1.6 GHz |
| Cache | 25 MB | 30 MB | 35 MB | 60 MB | 30.5 MB L2 | 34 MB L2 |
| Memory Speed / MCDRAM | 68 GB/s | 68 GB/s | 76.8 GB/s | 85 (60) GB/s | 352 GB/s | 115.2 / 448 GB/s |
| Thermal Design Power (TDP) | 105 W | 120 W | 135 W | 165 W | 300 W | 215 W |
| Memory installed | 64 GB DDR4 2133 MT/s | 128 GB DDR4 2133 MT/s | 128 GB DDR4 2400 MT/s | 4x48x16 GB DDR4 2133 MT/s | GDDR5 16 GB | 16 GB |
| Latest supported ISA extension (include previous ones) | AVX2 | AVX2 | AVX2 | AVX2 | IMCI | AVX-512 |

*Table 2 - Main characteristics of the different systems used for the architecture and instruction set comparison.*

#### 4.3.3.1 System level tuning

System tuning using Turbo frequency (TUR), Transparent Huge Page (THP), memory allocator (MAP) can be done without modifying the source code. This brings both performance gains as shown in Figure 10 and interesting information on dwarf behaviour. Indeed, enabling turbo offers a gain equal to 11%, enabling THP gives 22%, MAP 27%, and finally the best performance (35% of performance gain) is achieved by the combination of MAP and TUR. This inform us on the fact that memory management is a key point.

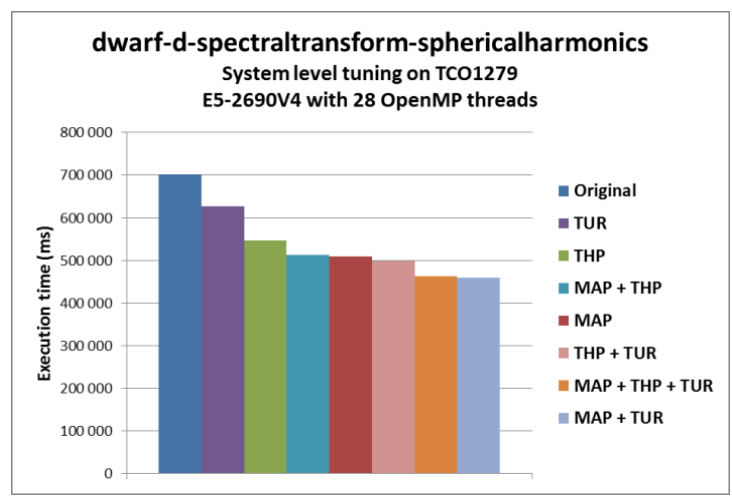

*Figure 10 - System level tuning performance results*

### 4.3.3.2 Intra node scalability study: architecture and instruction set comparison

The execution time according to the number of cores for 11 different configurations is presented Figure 11 (zoom Figure 12). It clearly shows that the last generation of Xeon (E5-2690V4) gives the highest performance while the KNC the lowest. Figure 12 highlights the gap when using hyper threading on the KNL and the SMP system.

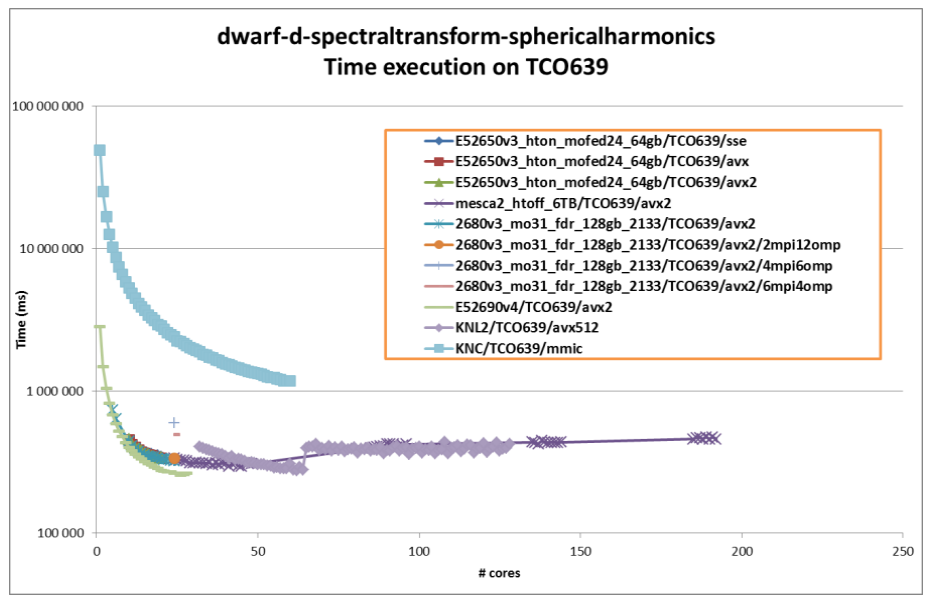

*Figure 11 - Spherical harmonics spectral transform dwarf intra-node scalability on different processors and configurations.*

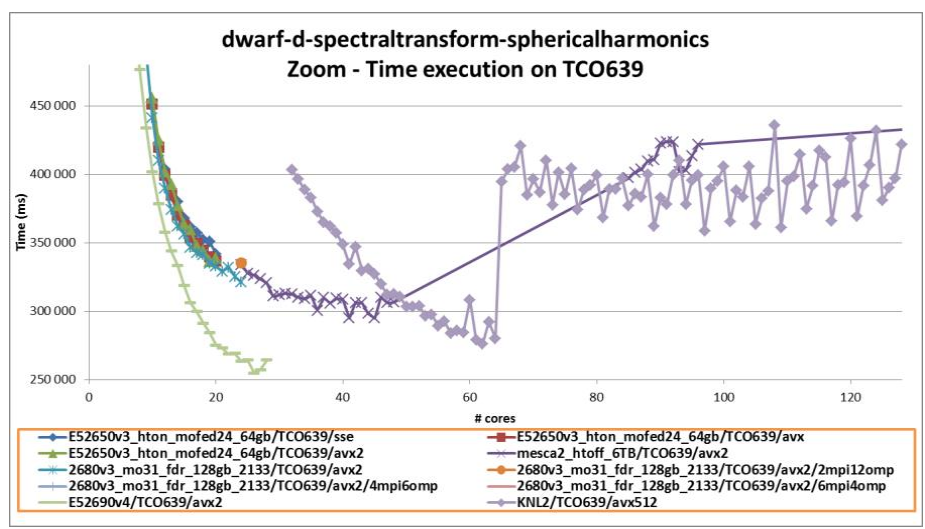

*Figure 12 - Spherical harmonics spectral transform dwarf intra-node scalability on different processors and configurations – Zoom.*

### 4.3.3.3 Directive based optimisation

According to the profiling results given by Intel Advisor (Figure 13), it appeared that 35% of the application is vectorized and as said before most of the execution time is passed in the MKL which is already optimized. The idea was to guide the compiler to vectorize more loops. This has been done by adding SIMD pragma (Optim 2 to 3 in Figure 14) but execution time has not been improved due to slow memory access. Adding contiguous pragma to the definition of some tables (Optim 4) leads to a performance improvement up to 11% on a Xeon processor, but the obtained gain is reduced according to the number of cores used. This is mainly due to the fact that the pipeline is stalled due to demand load or store instructions during a parallel execution. This shows that this dwarf is memory bound.

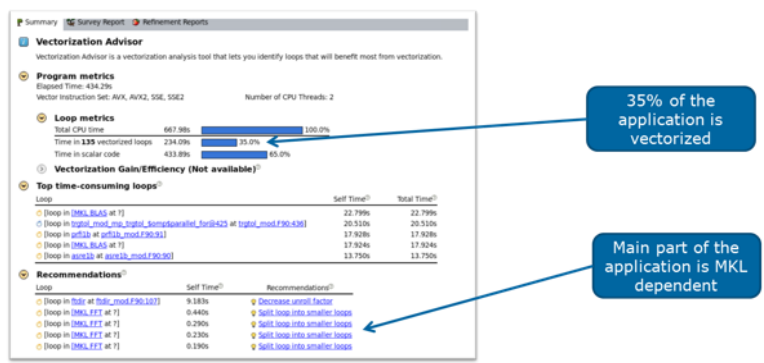

*Figure 13 - Intel Advisor profiling results*

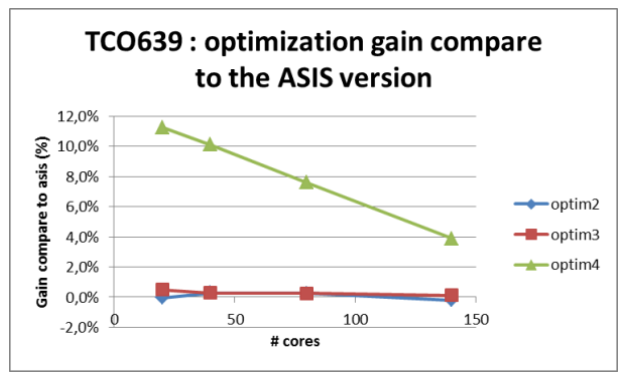

*Figure 14 - Optimization results using directive based approach*

### 4.3.4 Last obtained performance results

In order to propose a performance comparison, some experiments have been done recently on a high-end system node equipped with a dual-socket Intel Xeon Gold 6150 processors (2 x 18 physical cores running at a base frequency of 2.7 GHz). In this system, the best timing obtained to compute a time step is equal to 27.9 ms using OpenMP threads. The roofline graph from Intel advisor tool for this platform running the entire dwarf (100 time steps) with 36 OpenMP threads (1 per physical core) is given in Figure 15. The graph on the right shows MKL functions and loops as additional green points which are closest to the roof compared to non MKL part of the code, exceeding for some of them the L3 bandwidth (722.8 GB/sec), and reaching near 1 TFlops for some others.

### 4.3.1 Summary

In this part, a detailed profiling and benchmarking of the spectral transform – spherical harmonic dwarf has been presented in addition with a set of non-intrusive optimizations and their corresponding results. First of all, an intra-node scalability study was presented showing that the Atlas version with OpenMP parallelization on the Xeon 2690V4 gives the best performance. It also allowed identifying the best ISA to use, the best system level tuning (frequency and memory management policy). Finally, directive based optimization provided a gain of 11%.

The main outcome resides in the fact that this dwarf already uses an optimized library for the main kernels (FFT and MatMul) and is memory bound. As a conclusion, a better memory management can lead to a higher vectorization and batching the execution of the main kernels can intensify the arithmetic intensity to take advantage of the wide AVX2/AVX-512 registers. The next optimization strategies therefore imply a deep refactoring of the code, including memory structures, to achieve better performance which is future work.

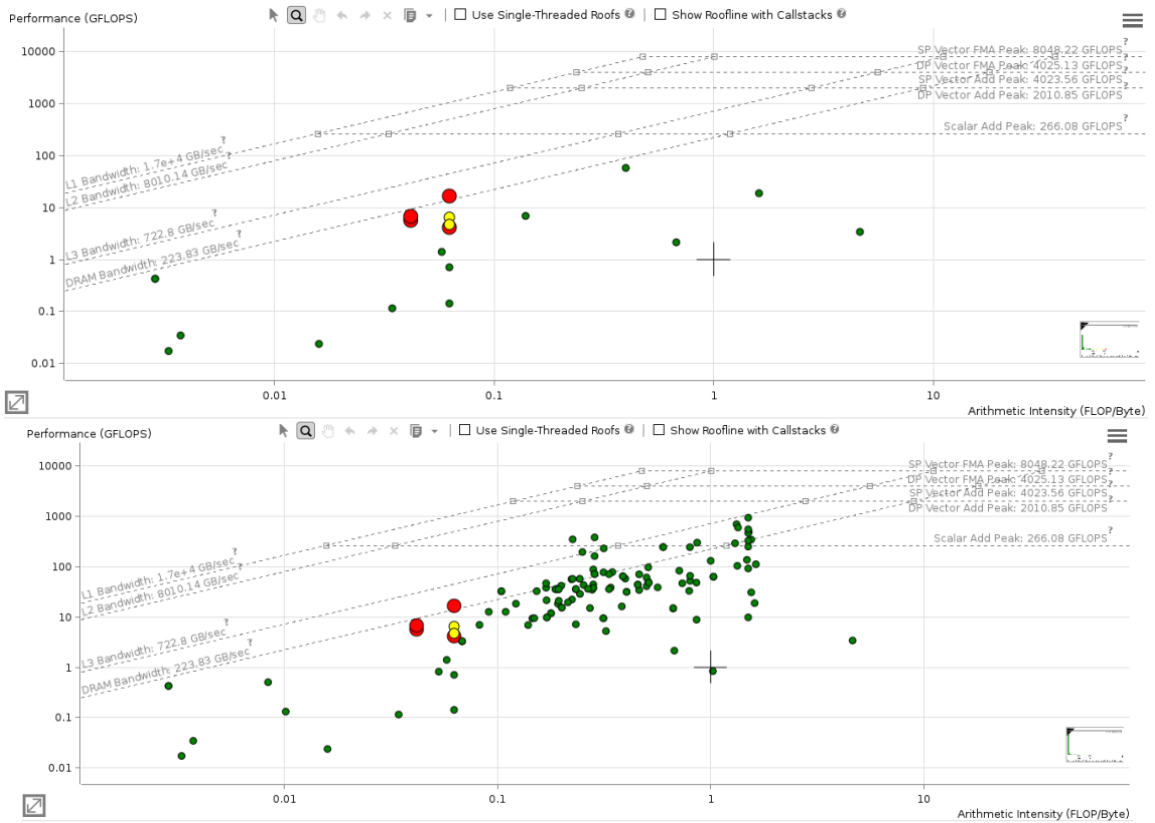

*Figure 15 – Spherical harmonics spectral transform dwarf roofline graphs without (upper) and with (lower) MKL loops and functions.*

### 4.4 Optalysys Optical Processor

#### 4.4.1 Introduction

The Spherical Harmonics dwarf is a fundamental part of the IFS (Integrated Forecast System). It represents a significant portion of the model processing time. Moreover, the all-to-all communication required in the 2D Fourier transform introduces a possible scalability bottleneck on conventional processors. Optalysys have been investigating an optical implementation of this dwarf which escapes this scaling regime. While there are a number of engineering challenges, this could offer a significant performance increase.

In this chapter, we discuss how the spherical harmonics transform can be implemented optically. Two methods are presented, with increasing optical sophistication and potential performance. Finally, we discuss the prospects of this technology.

Derivatives can be evaluated very efficiently and accurately in a spectral domain, where they become a simple pointwise multiplication. Such an approach relies on a spectral transform. For a global climate model – where the dynamics occur on the surface of a sphere – an appropriate spectral transform is the spherical harmonics transform. This transform is summarised in the following figure.

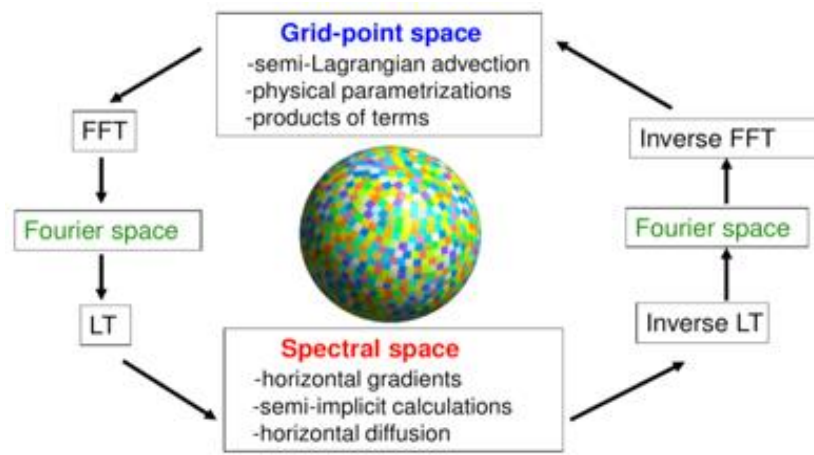

*An outline of the spherical harmonics transform used in the IFS. It consists of 1D FFTs applied along the longitudes, and a 1D Legendre transform applied along the latitudes.*

*(Reproduced from the ECMWF SphericalHarmonics dwarf documentation)*

We will consider in this document the 'forward' transform: from grid-point to spectral space, and implement it in this direction.

The vertical levels are denoted $l$, and $m$ and $n$ are the Fourier and Legendre coefficients respectively.

$$\zeta_l(\theta, \lambda) = \underbrace{\sum_{m=-M}^{M} e^{im\lambda}}_{\substack{\text{Fourier transform across} \\ \text{longitudes}}} \underbrace{\sum_{n=|m|}^{N(m)} \zeta_{n,l}^m \overline{P_n^m}[\cos(\theta)]}_{\substack{\text{Spherical harmonics transform} \\ \text{across latitudes}}}$$

### 4.4.1.1 Coherent optical information processing

We make use of some classic systems from coherent optical information processing. These systems rely on the fact that:

- A propagating optical field can carry 2D complex data.

- A simple lens can perform a complex Fourier transform on this data.

- An 'optical correlator,' which exploits the optical Fourier transform and the convolution theorem to produce a system capable of extremely high-performance pattern-matching tasks.

We will discuss each of these elements in turn.

### 4.4.1.2 Carrying information in a light beam

Consider a propagating laser beam. It is coherent; it has one wavelength and the waves rise and fall in a predictable sinusoidal pattern. We can consider a slice perpendicular to this beam. Light propagates in a deterministic manner. If we can fully specify the beam in one slice, we can define the beam along its entire path.

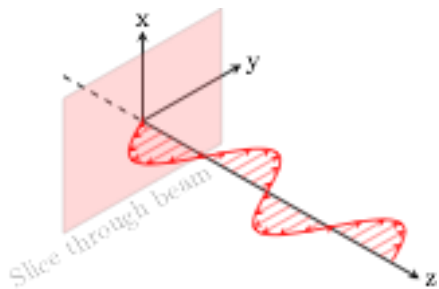

In order to define the beam across this slice, we need to specify how bright the beam is, and the phase of the sinusoid (at what point in the cycle we are at). This can be done using a simple complex number: the magnitude corresponds to the brightness of the laser beam and the phase corresponds to the point in the beam's oscillatory cycle.

Hence, a propagating coherent optical beam can be defined at one point in the beam by a complex function $A(x,y)$. We can modify this beam to encode information.

### 4.4.1.3 A lens performs a Fourier transform

What happens when a light beam encounters a lens? If we had a completely flat beam, we know it would get focussed to a single point. However, what if our beam was carrying information? What does the focal plane look like then?

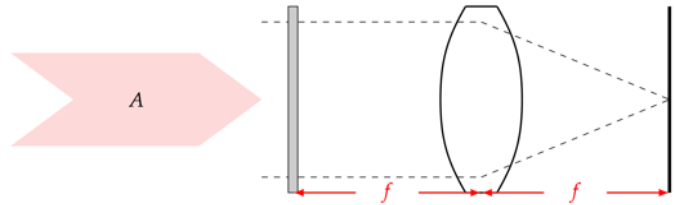

The focal plane contains precisely the Fourier transform – both amplitude and phase – of the complex field found at the back focal plane. This is consistent with the fact that for a perfectly flat beam, we get a single spot (the 'DC' term in Fourier theory).

Information can be input into the optical system at the back focal plane. For example, if a beam of complex amplitude $A(x,y)$ passes through a transparency with a spatially-varying transmission function $t(x,y)$, the resulting beam has an amplitude $A(x,y) \cdot t(x,y)$.

This optical Fourier transform is exploited in many applications. It forms the basis of many holographic techniques and underpins coherent optical processing. This powerful phenomenon is the driving force behind the Optalysys optical processor.

### 4.4.1.4 Encoding information in a light beam

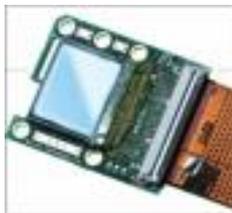

Information is encoded into the optical beam by using spatial light modulators (SLMs). These are essentially very small displays (indeed, some of the devices used in the Optalysys processor system originate in display projectors). These devices use liquid crystal technology (combined with linear polarisers) to modulate the light beam. In general, the magnitude and relative phase of the light beams can be modulated.

Each pixel of the SLM is addressable with an 8-bit value (256 levels). The SLM is not capable of independently modulating the magnitude and phase of the optical field. In the Optalysys processing system, the SLMs are configured to modulate both the amplitude and phase in a coupled manner, such that optimal correlation performance is achieved. A representation of a typical operating curve is shown in the adjacent Argand diagram.

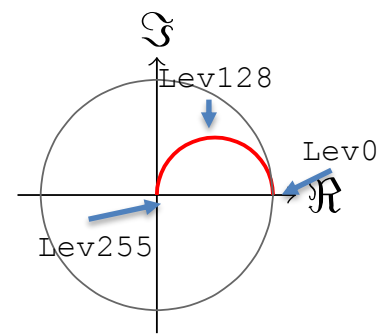

*An Argand diagram of the complex plane showing a curve where the 256 modulating levels offered by the SLM typically lie.*

### 4.4.1.5   The Optical Correlator

The Optalysys processor is a high-performance optical correlator. The optical correlator is the classic application of coherent optical processing. It consists of two lenses performing sequential optical Fourier transforms ($\mathcal{F}$), with a multiplicative filter situated at the intermediate focal – or Fourier – plane.

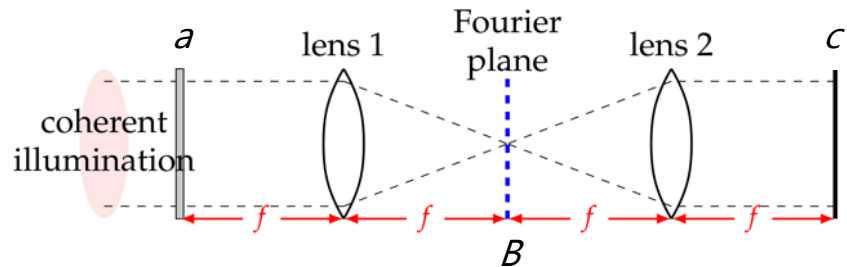

The filter effectively multiplies the optical field by a 2D function $B$. Input data $a$ are placed at the front of the system, and a camera sensor images the output beam $c$.[1] The system performs the mathematical operation

$$c = \mathcal{F}^{-1}\{ \mathcal{F}\{a\} \cdot B \}$$

where $c$ is a complex amplitude function. The lens in reality performs a forward, rather than inverse Fourier transform, but the net effect is a coordinate inversion compensated for by the camera sensor orientation. The camera sensor measures the intensity of this field,

$$I = |c|^2.$$

The convolution theorem uses the Fourier transform to effect the convolution (∗) of two functions, $f$ and $g$, by simple multiplication:

$$\mathcal{F}\{ f * g \} = \mathcal{F}\{f\} \cdot \mathcal{F}\{g\}.$$

By inspection, it can be seen that the effect of the optical system is to evaluate the convolution

$$a * \mathcal{F}^{-1}\{B\} = a * b.$$

One of the inputs to the correlation, $a$, is directly input into the optical system. The other input to the correlation, $b$, is derived digitally. This is the slow part of the process, and done off-line, producing $B$ using a digital discrete Fourier transform. (This fact indicates appropriate use of an optical correlator; there is some overhead in generating the filter $B$ from the target $b$). Optalysys excel in this filter design process and can provide appropriate assistance.

The optical Fourier transform and all of the functions are inherently two dimensional. The propagating light beam can be thought of as a 2D function propagating and transforming along a third direction. The system is most naturally applied to 2D datasets, and many problems can be mapped to an appropriate representation.

---

[1] Although at the correlation plane it's technically an 'imager,' not a 'camera,' in this overview we'll use the looser term 'camera' for its familiarity

### 4.4.1.6 Correlation

While it is the convolution theorem that fundamentally underpins the *4f* system, they are referred to as optical correlators. Correlation is tightly related to the convolution process, simply corresponding to reversing coordinates in the function being convolved with. Note that in 2D images, a reversal in each coordinate is a rotation of the image. As is customary in the literature of the discipline, we will define these functions in 1D for clarity, though they naturally extend to 2D.

**Convolution** (∗) of two functions *f(x)* and *g(x)* is defined, in both discrete and continuous representations, as:

$$f * g \ (x) = \Sigma_i \ f(i) \ g(x\text{-}i),$$

$$f * g \ (x) = \int f(\chi) \ g(x\text{-} \chi) \ d\chi.$$

**Correlation** (○) of the same two functions *f(x)* and *g(x)* is defined as:

$$f \circ g \ (x) = \Sigma_i \ f(i) \ g(x\text{+}i),$$

$$f \circ g \ (x) = \int f(\chi) \ g(x\text{+} \chi) \ d\chi.$$

From these definitions, it is clear that the operations are interchangeable under coordinate reversal of one of the functions. This reversal is done in the optical correlator simply by rotating the filter. For symmetric functions, correlations and convolutions are equivalent.

The optical Fourier transform is a true continuous Fourier transform. Consequently, the convolutions and correlation operations are, in principle, continuous. However, the optical system inputs and outputs are pixilated, meaning the data is discretised. While the operations can be considered to be discrete, the camera sensor pixels at the output do not sample at a single location (Dirac-function sampling), but integrate over a finite region.

Correlation is, amongst other things, very useful for pattern matching applications. The process is essentially dragging one function over another, and taking the dot-product between them at the set of all displacements. Two functions will have a large dot-product and produce an optical 'correlation spot' at locations corresponding to where their displaced versions match.

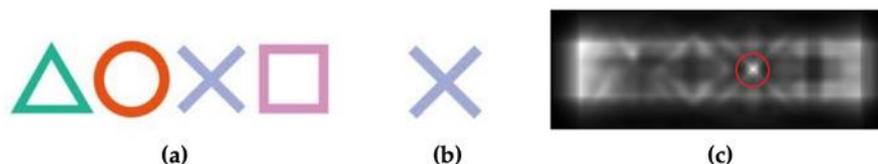

*An example of correlation of two functions. A test image (a) is correlated with a target (b). A peak in the output (c) shows where there is a match (highlighted).*

In an optical correlator, the first input, *a,* contains a representation of one function, while *B* is the Fourier transform of the other function. This transform is evaluated digitally, producing an appropriate filter (a frequency-domain representation of *b*). As will be discussed, this frequency-domain representation is not simply the Fourier transform of *b,* due to the characteristics of the devices used to encode information optically.

### 4.4.2 Extracting Fourier-Legendre coefficients using an optical correlator

The objective of the spherical harmonics transform is to extract the Fourier-Legendre coefficients, as determined by the definition:

$$\zeta_l(\theta, \lambda) = \underbrace{\sum_{m=-M}^{M} e^{im\lambda}}_{\substack{Fourier\ transform\ across \\ longitudes}} \underbrace{\sum_{n=|m|}^{N(m)} \zeta_{n,l}^m \overline{P_n^m}[\cos(\theta)]}_{\substack{Spherical\ harmonics\ transform \\ across\ latitudes}}$$

Our objective is to populate a 2D spectral grid with these coefficients (at each vertical level *l*). Ideally, we would perform this in one optical operation (as each output point depends on all input points, all of the input data must be present), but there is no straightforward optical system which performs this transformation.

Note that each coefficient is found essentially by taking an inner product. We are projecting our input data onto an appropriate Fourier-Legendre basis function. We can evaluate this projection simply by taking a dot product.

We have at our disposal an optical correlator. As previously discussed, an optical correlator essentially evaluates dot-products (the convolution that the system performs is essentially shift-invariant dot-products). We can use an optical correlator to serially perform the dot-products and extract the coefficients.

#### 4.4.2.1   Using a *4f* optical correlator to extract Fourier-Legendre coefficients

Recall that an optical correlator performs the operation:

$$f \circ g\ (x) = \Sigma_i\ f(i)\ g(x+i),$$

Where *f* is a function which is directly input to the optical domain and *g* is the Fourier transform of a filter function *G* applied in the Fourier plane.

At *x=0*, we are evaluating exactly the dot-product between *f* and *g*. Hence if *f* is our 2D function in grid-point space, and *g* is a Fourier-Legendre basis function, at *x=0* we extract the correct projection.

Of course, the optical correlator is evaluating the corresponding shifted dot-products at all other values of *x*. Unfortunately, we cannot make use of these as they do not mean anything in the context of our particular problem here. This may seem somewhat decadent – we are evaluating a full convolution for the sake of finding one dot-product – but as we are doing this optically we essentially get it at no extra cost. In order to extract different coefficients, we need to change the function *g*, which we can accomplish by cycling through different filters.

The process for extracting one coefficient is as follows. An input area of data – 'grid-point space' – is presented on the first SLM ($f$). We want to take the dot product with a specific Fourier-Legendre basis function ($g$). This function is converted into a filter function ($G$) (discussed later) and shown on the filter SLM. A camera samples the output plane and the *(x,y)=(0,0)* point gives the magnitude squared of the dot product. This process is summarised for one of the Fourier-Legendre basis functions below.

### 4.4.2.2 Experimental demonstration

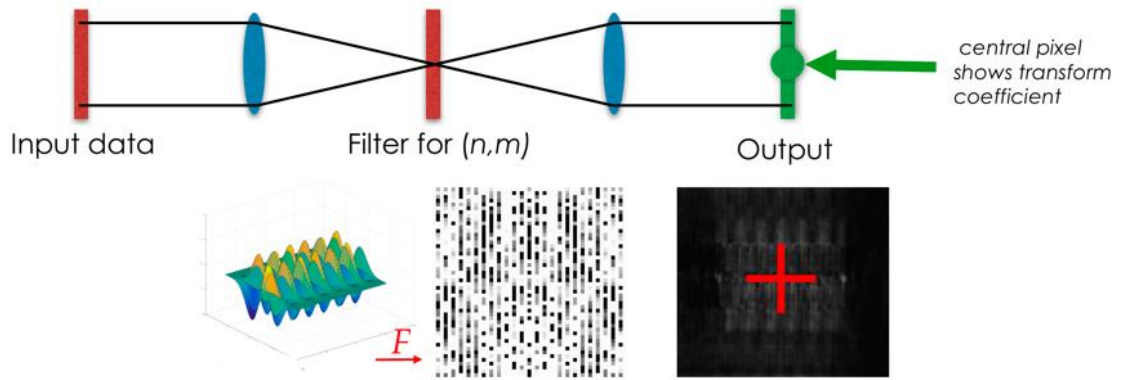

*Using a 4f correlator to measure Fourier-Legendre coefficients*

*An input area of data – 'grid-point space' is presented on the first SLM. We want to take the dot product with a specific Fourier-Legndre basis function (inset, here (m,n) = (5,6)).*

*We convert this basis function to a corresponding filter using an FFT. The output then contains the convolution of the input data with this basis function. The central intensity of this convolution is the dot-product we are interested in.*

To demonstrate this technique, we present the case of extracting one Fourier-Legendre coefficient from a dataset (the (5,6) coefficient shown in the previous figure). We start off taking the auto-correlation and then perturb this by different amounts, up to completely modifying it so that there is no projection onto the selected basis function. We perform this experiment over many different randomly perturbed functions, and plot the results as a histogram.

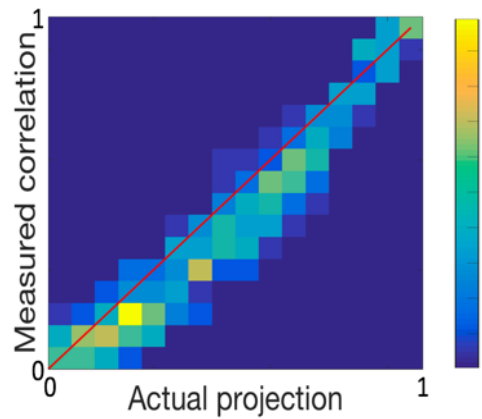

*Optical measurements of the $(m,n)=(5,6)$ Fourier-Legendre coefficient, using different randomly-perturbed functions. The optical intensity measurement is up the $y$ axis; the correct computed value is along the $x$ axis.*

*Note that these values are of the optical intensity (i.e. the computed value is the modulus-square) of the Fourier-Legendre function).*

It is clear that we are having some experimental success here. However, it should be noted that there is clearly a systematic error (likely due to the sampling size of the camera sensor being finite, rather than a true point-like Dirac sampling). This error will be exaggerated further in amplitude, rather than intensity. Furthermore, as will be discussed in the section on filter design, there are accuracy implications due to compromises in filter implementation.

### 4.4.2.3 Parallelising use of the filter function

Such a system is relatively slow (it can only extract one coefficient per frame) as, due to the liquid crystals used, the cycle time of the system is low. Furthermore, in using a camera we are massively over-provisioning our output bandwidth (we only use one pixel each frame).

However, there is one trivial way in which the process can be sped up. If there is excess resolution on the input SLM, this can be used to tile input data. For each input tile, we can extract the same Fourier-Legendre coefficient in parallel. This would be useful, for example, to parallelise the spectral transform across altitude level *l*.

In doing this, for each tiled input we are basically shifting our origin. Due to the shift invariance of the correlation operation, this means that at the output we simply have to correspondingly shift our sampling point to find the correct dot-product. An example of the output seen when such an approach is undertaken is shown in the following figure.

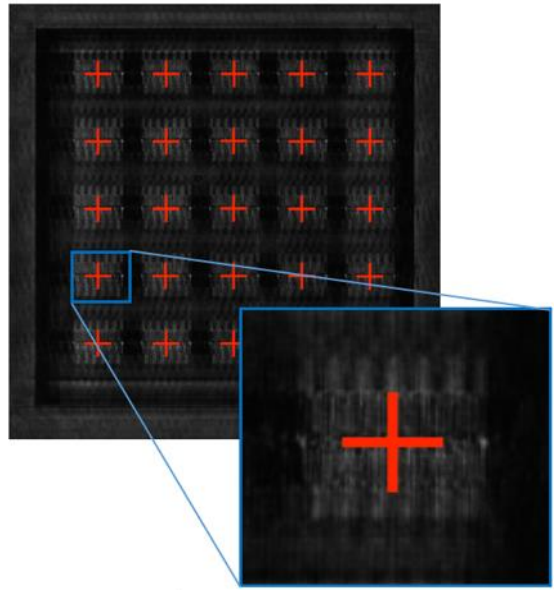

*By tiling input functions, we can parallelise the output of a given Fourier-Legendre coefficient across different input datasets. Shown is the optical output when such a process is undertaken. The red crosses indicate the appropriate sampling points to find the dot products.*

Parallelising the filter function has an effect on the competiveness of the optical approach relative to a digital approach. While in this 'toy example' we are not demonstrating a performant application, this is broadly relevant.

The attraction of an optical approach is that it offers a Fourier transform as an *O(1)* process, as compared to an *O(nlogn)* process as offered by a digital Fourier transform. The size of the Fourier transform is fixed by the resolution of the system. It is when exploiting this resolution that we can achieve the biggest performance gain.

When we do not make use of the full resolution Fourier transform, we do not make use of the full system performance. In an optical correlator we can still make use of the full resolution by tiling different inputs together at the input, and separating these inputs at the output, as shown in the case above.

While we are still making use of the full resolution of the system, we are not capitalising on the high-resolution of the corresponding Fourier transform. This 'batch' approach is instead equivalent to a number of smaller Fourier transforms.

Our objective here is to consider the fraction of the latent system performance 'lost' when we do not make use of the resolution of the inherent Fourier transform, but instead use it to perform a batch of lower-resolution transforms. We will make this comparison using simple computational scaling arguments.

We will consider the comparative computational scaling as being dominated by the scaling of the Fourier transforms (the elementwise multiplication is 'free' by comparison). Thus, we can compare our performance respectively as:

$$FFT: \quad O(N \log N)$$

$$OFT: \quad O(1).$$

Where we have *N* pixels in total. However, what if instead of evaluating one size-*N* transform, we are evaluating *P* size-*M* transforms, where *N=M·P*. We still have the same *O(1)* scaling for the OFT, but our FFT now has scaling:

Batch-FFT:  O(P·M log M).

We can define a 'scaling slowdown,' *S*, which describes the factor of the roofline performance we can achieve, subject to these simple scaling arguments.

$$S = \frac{P \cdot M log\, M}{N\, log\, N} = \frac{P \cdot \frac{N}{P}\, log\, \frac{N}{P}}{N\, log\, N} = \frac{log\, N - log\, P}{log\, N}$$

$$S = 1 - \frac{log\, \mathrm{P}}{log\, \mathrm{N}}$$

(Note that it does not matter the bases of the logarithms in this formula). To recap, this formula is the fraction of the roofline performance advantage of the optical system relative to a computer we expect to realise when performing batched operations.

For a 10 MPx (10E6) pixel input, we can plot this slow-down factor:

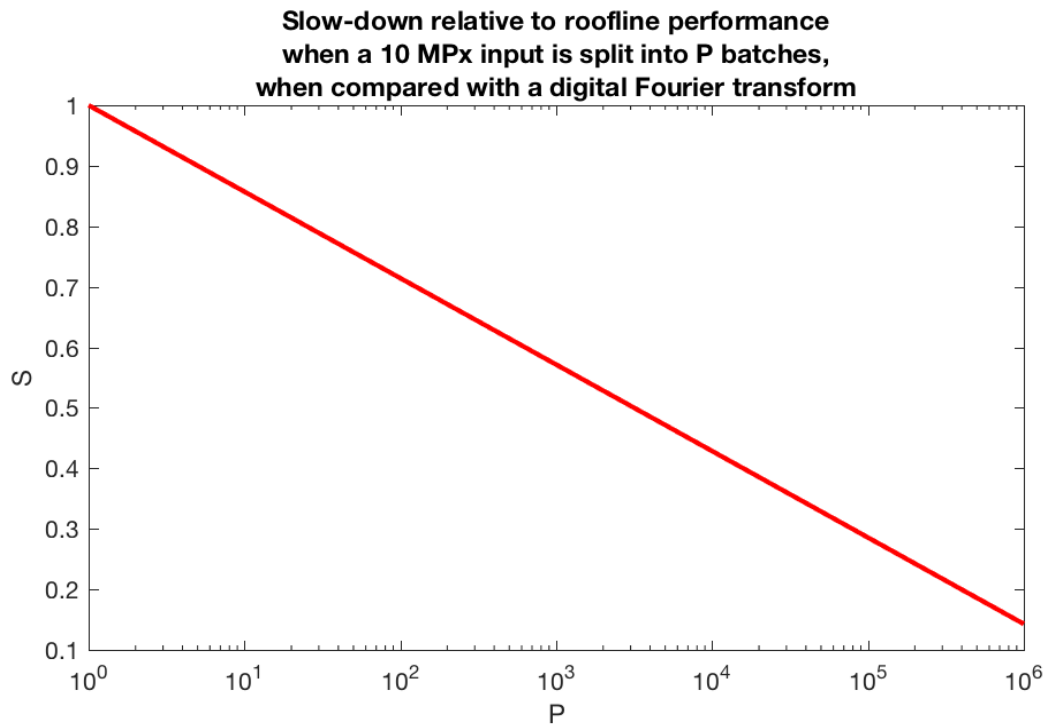

### 4.4.3 A multichannel astigmatic optical processor

The extraction of the Fourier-Legendre coefficients using the *4f* optical correlator is a robust implementation using a classic optical architecture. However, it suffers significantly in terms of performance.

While this optical implementation has a favourable scaling region of *O(MN)* where *M* represents the Fourier transform coefficients across longitudes, and *N* represents the Legendre transform coefficients across latitudes, in reality we would expect performance to be somewhat lacking.

A method is required to make better use of the optical capabilities. An obvious step is to take advantage of the fact that the Fourier transform is the operational primitive of coherent optical information processing. It is obtuse that in the *4f* correlator implementation described in the previous section we are implementing the Fourier transforms with a correlation filter when optics so naturally implements this operation.

To this end, we have developed a system which combines the fact that coherent optics provides a natural platform to evaluate Fourier transforms with the ability of an optical correlator to evaluate Legendre coefficients. In essence, this system cascades these two operations. Recall that – in the forward direction of this transform – we are looking to perform a Fourier transform in one direction, followed by a Legendre transform in an orthogonal direction.

### 4.4.4 Performing parallel 1D Fourier transforms

On order to perform parallel 1D Fourier transforms, we use an astigmatic (axially asymmetric) system. Specifically, we make use of cylindrical rather than spherical lenses. Cylindrical lenses exert optical power along one direction, but not along the orthogonal direction.

A system which performs a 1D Fourier transform on a complex field of optical data is shown below:

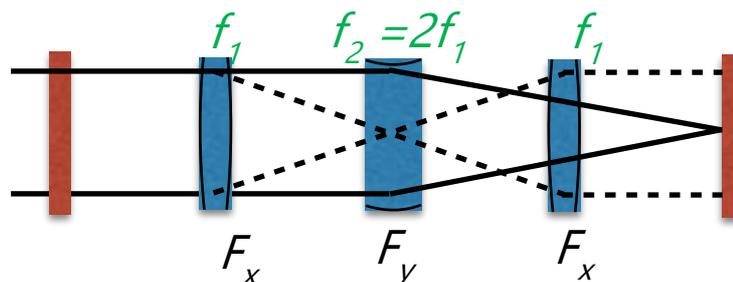

*An astigmatic system that performs a 1D Fourier transform (in the y-direction) from input to output using cylindrical lenses. The paired shorter-focal-length lenses act as an imaging relay in the x direction, while the longer-focal-length lens performs an optical Fourier transform in the y direction.*

This system essentially consists of an imaging system in one direction (*x*) and a Fourier-transforming system in the other direction (*y*).

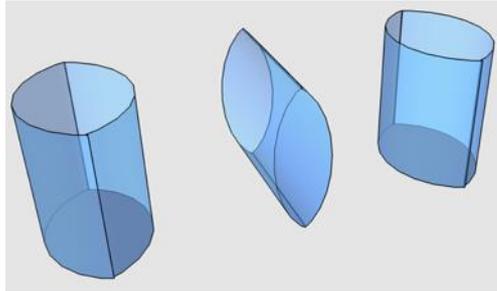

*A 3D rendering of the cylindrical lenses required to perform a 1D Fourier transform*

### 4.4.5   Performing parallel 1D Legendre transforms

We wish to use a similar approach to the previous system whereby we use an optical correlator to find the projections of functions onto different basis functions, effectively by evaluating a dot-product. This system in principle requires a 1D optical correlator. We could use two of the 1D OFT assembles shown in the previous section with a corresponding intervening filter to achieve this.

However, there is an optically more straightforward approach. Even though we wish to perform 1D correlations, we can use spherical lenses and a classic 2D optical correlator. In order to evaluate 1D operations, we simply need to ensure that the filter is uniform in the direction we wish to preserve. By doing this, we will not mix data in this direction and are able to implement a parallel processor.

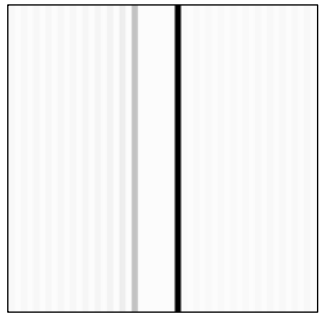

*A filter to extract Legendre coefficients in parallel (in this case n=3) in a 2D optical correlator. The uniformity in the vertical direction ensures that the system evaluates the coefficients for data arranged in the horizontal direction in parallel.*

### 4.4.6   Integrated optical design

We can cascade these two modules (a 1D Fourier transform stage and an optical correlator together):

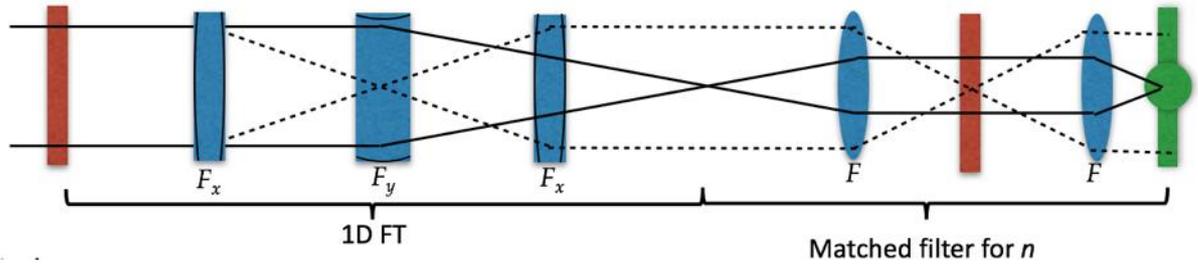

However, this is not the most straightforward optical design. Consider that from the input to the filter SLM (the two red planes), we are performing three Fourier transforms in the $x$ direction and two Fourier transforms in the $y$ direction. Neglecting inversions – which we can compensate for by rotating SLMs and cameras – this is equivalent to performing one Fourier transform and two Fourier transforms respectively. Hence, we can rationalize our optical system to:

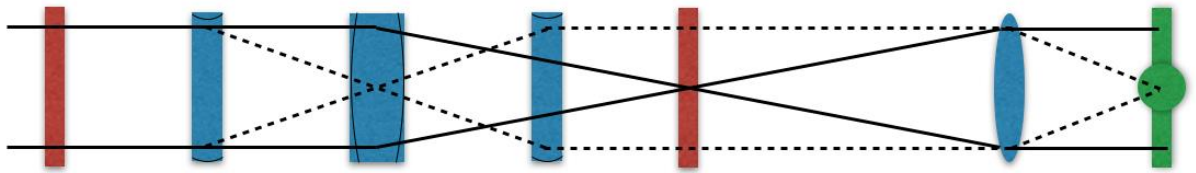

where we have rotated the 1D FT stage and omitted the first spherical lens from the optical correlator. This optical system performs the operations we require of it.

### 4.4.7 Implementation

We implement this system using commodity optical components and Sony 2K SLMs, which are found to have high optical quality, although unfortunately they have a shallow optical modulation depth of ~π.

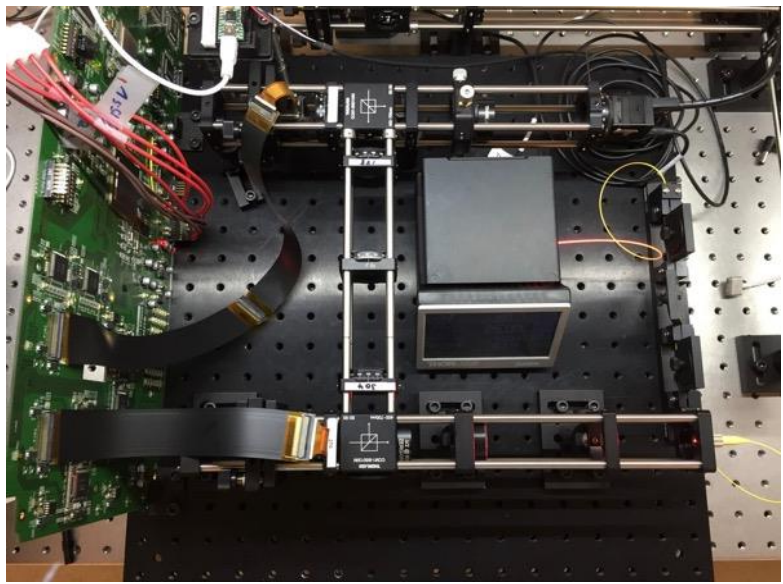

*The prototype multichannel astigmatic optical processor*

This is a challenging system to implement and represents in and of itself a novel optical architecture, with the associated challenges. In particular, due to the parallel nature of the convolution stage, the optical output is essentially dimmer than in a more conventional system as the contributing regions on the input plane are 1D lines rather than 2D patches.

Obtaining high-fidelity numerical results from this system is challenging and still a work-in-progress. However, we have managed to demonstrate the success of the fundamental architecture by successfully implementing a parallel 1D optical correlator.

It is first important to explain the geometry of the output plane. It has 2 axes. The first is the axis along which the 1D FFTs have been applied. We are interested in all of the values here, corresponding to the *m* coefficients. The second is the axis along which we are extracting the Legendre coefficients by using the centre of the correlation to evaluate the projection onto the basis function. We are only interested in these central values.

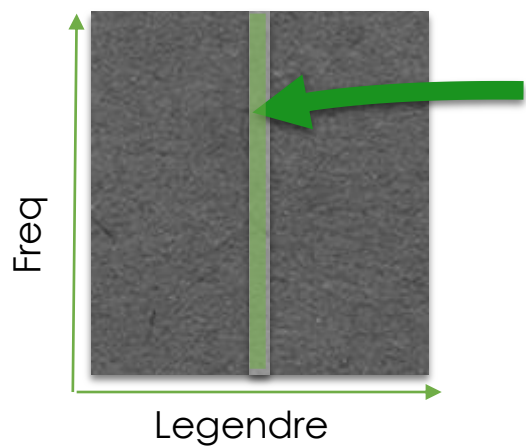

As such, an appropriate sensor for this application would be a 1D line sensor, rather than a conventional 2D camera sensor array. This is advantageous as such sensors offer very high framerate, meaning that camera bandwidth limitations would not be a problem.

The optical output when a Fourier-Legendre basis function is set as the input, with the corresponding Legendre function used as the filter shows the expected behaviour.

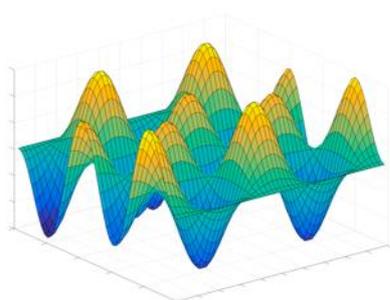

*(m,n) = (3,6)*

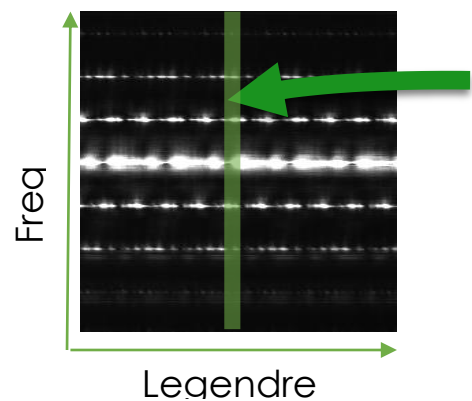

We can see that in the frequency direction we are seeing the different harmonics associated with this 3$^{rd}$-order function. In the Legendre direction, we see peaks corresponding to the appropriate auto-correlations (projections) lined up.

A more concrete demonstration is the presence of a correlation peak corresponding to a 1D input function, correlated with an appropriate corresponding 1D filter. The 1D input function is modulated in the orthogonal direction by a simple sine function to carry the corresponding data out into harmonics at the output plane, by the effect of the 1D Fourier transform system front-end.

However, quantitatively extracting different Fourier-Legendre coefficients is challenging. There are a number of issues. For example, 1D functions represent small footprints, so light becomes a limiting factor. Assembling high-performance astigmatic systems is challenging due to the extra degrees of freedom associated with non-rotationally-symmetric components. In general, the tolerance on rotational alignment is particularly challenging. Nonetheless, this approach represents an exciting first-step

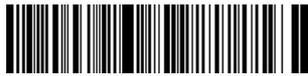

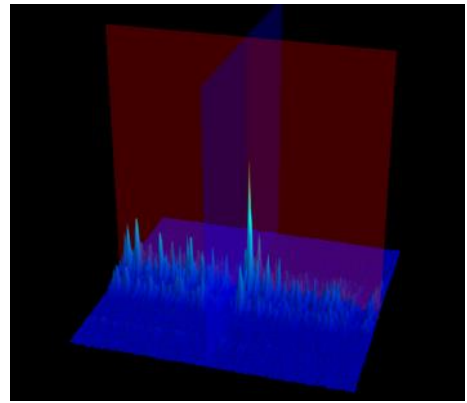

*A random 1D function used at the input. This is encoded by modulating it at a given frequency in the orthogonal direction and is displayed on the input SLM. This carries the information out into different harmonics along the frequency axis in the output plane. A corresponding 1D filter is displayed on the filter SLM and a correlation peak observed.*

towards an optical system which can evaluate this transform at the heart of NWP.

### 4.4.8 Prospects

The prospects of this approach are exciting. Fundamentally, this approach changes the scaling regime of the Fourier-Legendre transform dramatically. Current computational implementations are $O(N^3)$. By moving to an optical implantation, we can reduce this to $O(N)$ scaling.

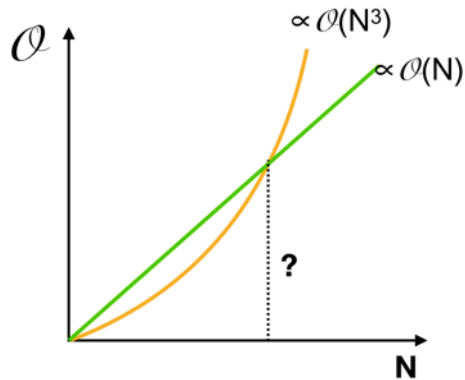

*Scaling behaviour of an O(N) and O(N³) process*

This is because the optics evaluates the Fourier transforms and Legendre projections with *O(1)* scaling, the *O(N)* comes from the fact that we have to iterate over different filters for each set of Legendre coefficients we wish to extract.

The issue is that, because we have to switch filters to extract the different Legendre coefficients, our system speed is limited by the refresh speed of currently available SLM technologies. For multilevel SLMs (which are preferable), we are constrained to ~100 Hz. Extracting 100 sets of Legendre coefficients in a second is too slow. If we could make this system work well using binary filters, we could achieve ~20,000 sets of Legendre coefficients per second, which is more viable.

A further factor to consider is how to extract either bipolar or complex coefficients. Optically, we measure the intensity. Depending on the direction of the transform being considered, the result may be either real or complex. Determining the relevant sign or phase will require techniques similar to those used to obtain the phase in the optical implementation of the biFFT dwarf. Namely, the symmetry of the operation will be exploited to constrain the result, with the response of the system to select perturbations used to determine these constrained values. There will be a computational overhead associated with this decomposition and reconstitution of the data.

### 4.4.9 Generating the filter functions

An ideal implementation of the filter would optically represent the Fourier-Legendre basis function *g* as a filter simply by taking the Fourier transform such that

$$G = \{g\}$$

However, we do not have a suitable modulator to be able to represent this fully complex function. We are restricted to the available action of the SLM: it's 'operating curve.' (So-called because the action of the SLM is described by a curve on the complex plane).

This hardware limitation is one of the main issues encountered in optical information processing (OIP). The academic literature is full of techniques to either: enhance the modulation range of SLMs; or to encode a given function effectively despite the compromises which must be made due to the limited modulation range. A hallmark of OIP is that successful systems can be implemented despite this limitation.

However, most OIP systems do not operate under the strict regime of numerical accuracy required by this application. The canonical OIP system is the optical correlator, used for optical pattern matching. In this application, the objective is to detect a given target within a larger data set. This does not require high numerical precision evaluation of a prescribed numerical function. Such an application can be designed to be robust to device shortcomings and imperfections.

This is not the case when high numerical precision is required. For the purpose of extracting a specific Legendre coefficient, there is one correct filter function *G*, and only one. Any deviation from this function represents a decrease in the numerical fidelity of the system. Some decrease in fidelity will be tolerable in any finite-precision system; but beyond a certain point this will have significant performance consequences.

### 4.4.10 An optimal method for creating the filter function

A number of methods are prevalent in OIP for dealing with the restricted modulation range of a given filter. However, an accepted – and provably optimal for the case of an optical correlator – method is to select the points on the operating curve of the device (as expressed on a complex plane) which minimise the Euclidian distance from the desired points [Juday93].

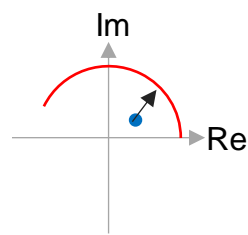

*The red line shows a potential operating curve (the range of accessible modulation states) for an SLM, as represented on the complex plane. The blue point is a target complex value we wish the filter to express. The point in the operating curve which minimizes the Euclidian distance from the target is the optimal choice.*

This is the approach which is adopted in this work.

The effect of this compromise has varying effects depending on the filter selected. Different target Fourier-Legendre functions suffer different detriments through this mapping process. For example, the restricted range of the operating curve means that for some filters, the Euclidian-distance minimising mapping is completely blank. Clearly this does not lead to satisfactory system operation!

However, this principle is only part of the story. The minimum-Euclidian distance principle tells us the optimal mapping given a particular operating curve, but it does not reflect the fact that we have some scope to manipulate this operating curve.

Firstly, the curve can be manipulated in terms of the hardware. It is a property of both the SLM selection and the polarisation scheme. By rotating polarisers or other polarisation control elements, we can change the polarisation environment and hence the operating curve.

Secondly, the operating curve can be rotated and scaled without affecting the operation of the system. Rotating and scaling in the complex plane correspond to multiplying by a scalar complex number, which does not have a meaningful effect on the operation of the system. Applying the minimum Euclidian distance principle combined with this scaling-and-rotation operation yields higher-performing filters.

### 4.4.11 Increasing system performance via in situ optimization

No optical system is perfectly constructed. Deviations from the ideal system are inevitable when the relevant length scale is the wavelength of light. Particularly relevant are variations in the flatness and modulation performance of the SLMs across the devices. There are two fundamental ways in which this can be dealt with:

1. Through characterisation of the devices, and compensating for deviations during operation of the device.

2. Through in situ optimisation of the device. Corrections are represented agnostically by some appropriate parameterisation, and these parameters are established through some in situ process.

Optalysys has experience with the former technique. While it does represent a 'gold standard' – in that the error that is being compensated for can be explicitly measured, and the optimal correction applied – it comes with significant disadvantages.

- It is time consuming and difficult to implement. Characterisation requires sophisticated interferometric setups.

- Characterisation data must be stored and associated not only with the physical device, but with the exact location and orientation of the device. While absolutely tractable, making devices non-interchangeable (and even explicitly not spatially invariant) comes with a significant overhead.

- Only things which are explicitly measured can be compensated for. This is challenging because some errors only manifest themselves actually in the specific system operation, and it can be very challenging to make appropriate measurements actually within the system.

The attractions of an *in-situ* optimisation technique are clear. Optalysys have developed such a technique, initially applied to the problem of optimising optical correlator filters. However, it could be generalised to the problem of improving the fidelity of optical numerical computation.

The general problem is to find a general algorithmic correction which can be applied during the filter design process, which compensates primarily for non-flatness of the SLM – which introduces phase variations and variability in the modulation performance across the device. However, by not being explicit we can also expect that this correction will also correct for other optical imperfections (focus, aberration).

For this exercise, we will consider the case of optimising a binary (2-level) filter in an optical correlation application. We are looking to maximise the height of correlation peaks when the input matches a given target.

The algorithm used when designing such a filter is:

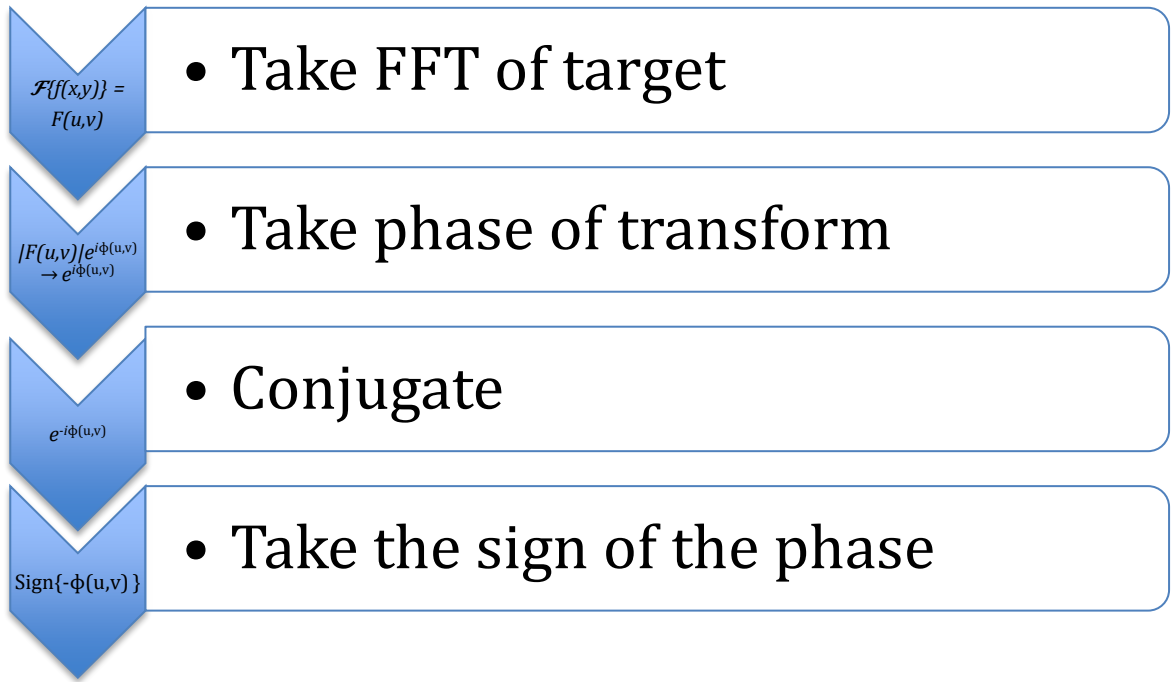

- Take FFT of target
- Take phase of transform
- Conjugate
- Take the sign of the phase

More information on this algorithm can be found in the literature on this topic. In some sense it is remarkable how well the phase only binary filter performs in a matched filtering application, but this is a well-known phenomenon in the field of optical correlation.

In principle this filter requires a binary phase SLM to display it – i.e. a panel which effectively represents the values {-1,1}.

Consider that we are using a multi-level SLM to display this filter. We have many possible ways we could map the filter onto the SLM levels. In general, the range of potential corrections that could be explored is too large to find the optimum filter.
It is at this point that we must use our physical intuition to constrain this parameter space.

We assume that any errors have an underlying physical explanation, and that they are likely to be smoothly varying. Hence, we parameterise these corrections using a set of smoothly-varying basis functions. A common choice in the world of optics is to use Zernike polynomials, which form a complete orthogonal basis on the unit circle. Furthermore, the different lower-order polynomials represent physically intuitive effects like tilt, defocus and low-order aberrations.

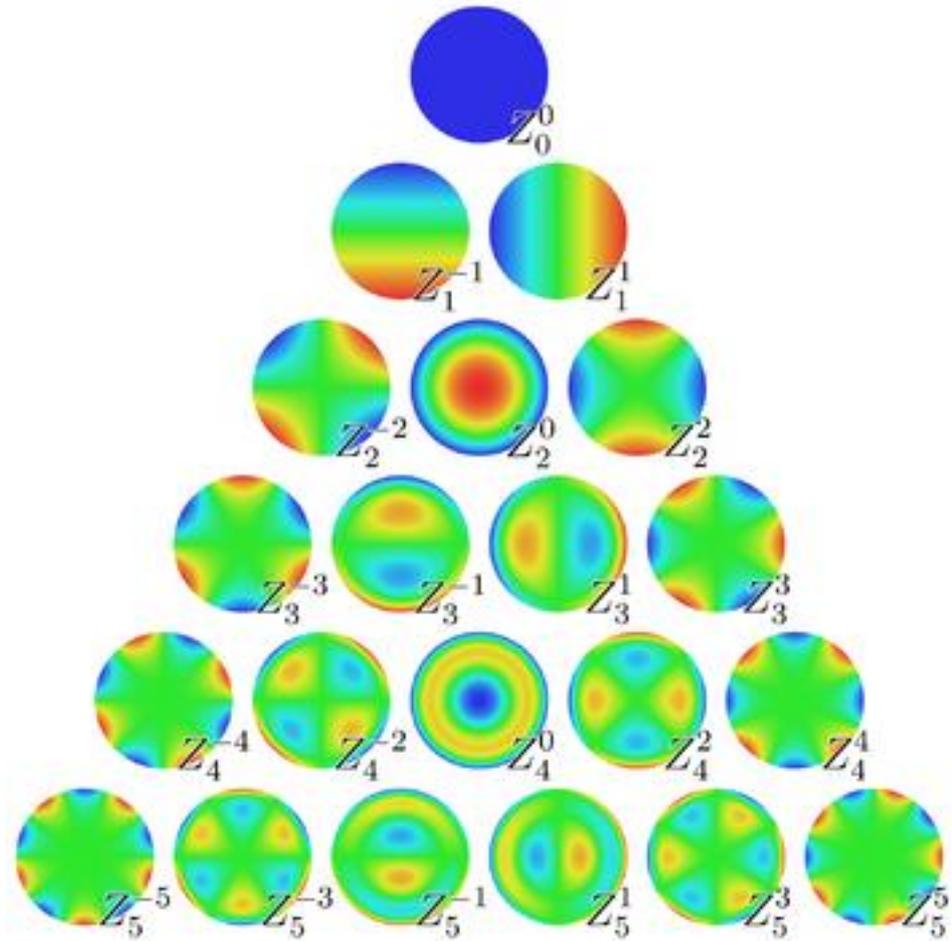

*The lower order Zernike polynomials. From https://en.wikipedia.org/wiki/Zernike_polynomials*

We select the first 6 polynomials to parameterise our correction functions.

We consider two correction functions applied to the previous algorithm. These are both phase functions. The first correction $\Phi_P$ is applied to the raw target phase of the function, which in principle compensates for non-flatness of the SLM. The second correction $\Phi_T$ is used to define a thresholding value which is not simply a phase of 0. (The fact that phase is mod($2\pi$) means that these corrections are not the same thing.)

An *in-situ* optimisation process across the parameterisations of these two functions is implemented, using the correlation peak height to measure when a good convergence is obtained. In order to ensure that we are not generating a correction which is only applicable to one filter, when determining the fitness function, we measure the peak height for the correction applied across a range of random targets (10 seems to be sufficient). Moreover, this method is found to be effective in different system configurations.

Plots of the resultant phase correction and threshold level corrections derived from the low-order optimised Zernike basis function coefficients are shown:

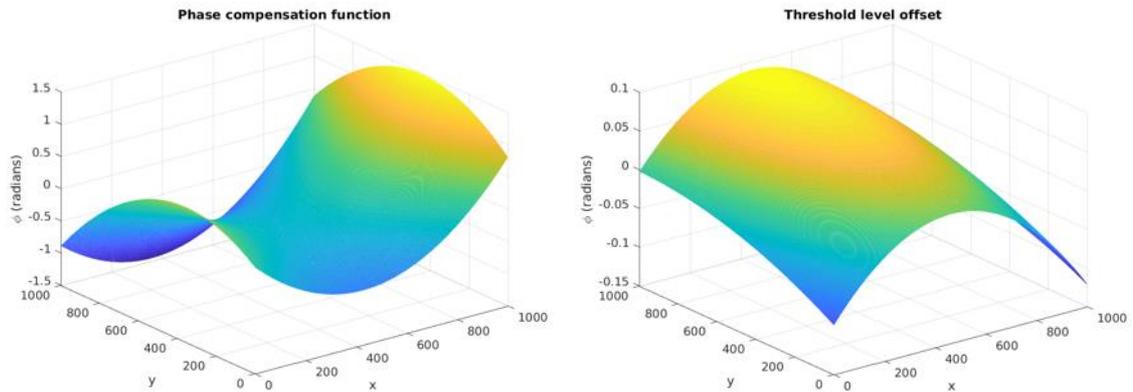

*The phase compensation function and threshold level offset obtained from the simulated annealing optimisation process.*

These correction functions are applied in the filter design process.

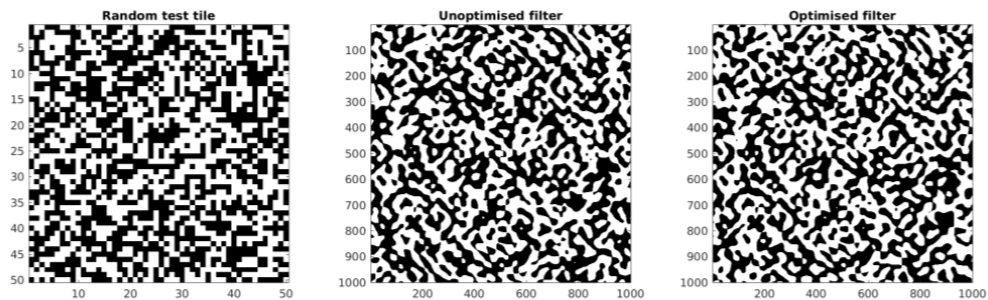

Left: *A random test tile which we wish to generate the filter for.* Middle: *The pre-optimised binary filter generated by the algorithm outlined above.* Right: *The optimised filter after the phase correction and threshold level offset have been applied.*

In order to assess the performance of this function, we obtain histograms of the peak heights for a number of randomly generated input and target pairs.

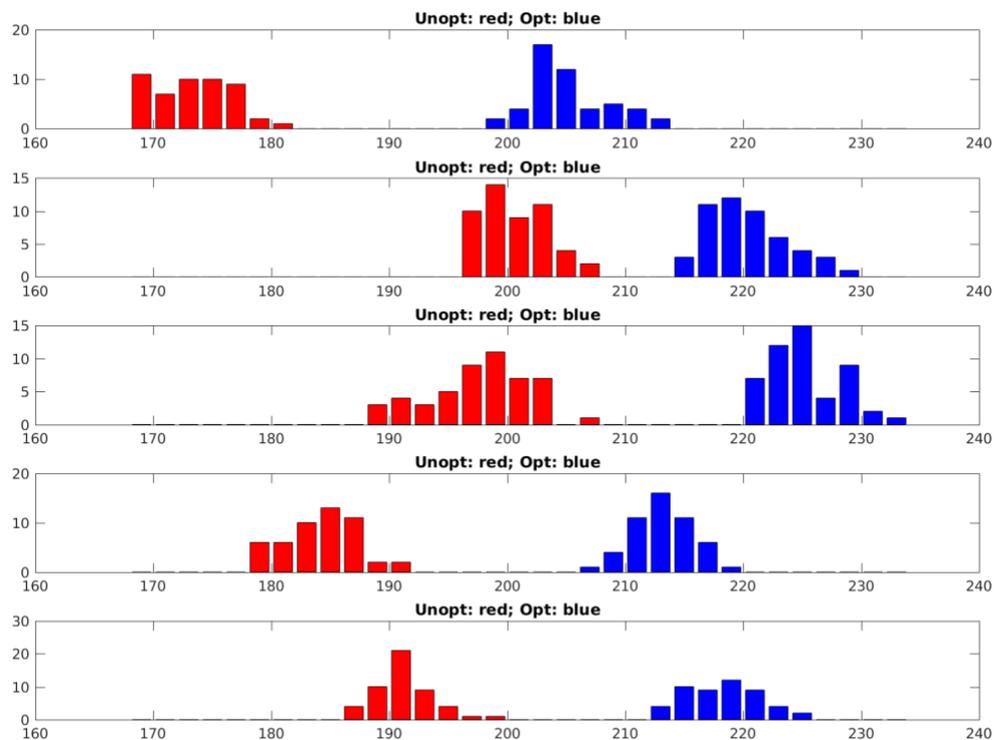

*For 5 different randomly generated test patches and filters, the pre-optimised and optimised peak heights when measured repeatedly. The effect of the optimisation process across these different test cases produces a clear improvement.*

There is a clear improvement in the correlations due to the optimisation process implemented. Furthermore, the corrections developed are generally applicable – we have not developed an optimisation that is only relevant to one filter.

In order to be more generally applied – for example, to numerical calculations – we need to generalise this approach. In particular, parameterisations which are applicable to multi-level filters need to be developed. The appropriate parameterisation is likely of geometric transforms to a specimen operating curve represented on the Argand plane.

Nonetheless, these results demonstrate the power of in situ optimisation techniques to improve the performance of an optical system.

### 4.4.12 Future hardware requirements of an optical processing system

There are a number of different hardware developments that are required to perform high performance, high fidelity numerical calculations. Compared to other applications of optical information processing, numerical weather prediction and associated numerical fields are uniquely challenging. The precision requirements from what is essentially an analogue system place high demands on the hardware.

The SLMs themselves in particular need further development, both in terms of the consistency and fidelity of the devices, and the throughput of them. The relatively slow speed of multi-level liquid crystal devices is an issue when trying to map numerical problems into a performant optical system. High speed SLMs are available, but they generally only modulate in two states. There are candidate technologies for high-speed multilevel devices (such as analogue ferroelectric liquid crystal devices, solid-state multiple quantum well modulators, and integrated photonic emitters), but they require further development into commercial products.

### 4.4.12.1 Complex modulation

A particular requirement, to which some progress has been made, is a method to faithfully implement the required filter functions. In general, this requires full complex modulation. Obtaining this is challenging, but not impossible. Doing so with high fidelity and at high resolution, however, represents a significant engineering challenge.

There are a number of potential routes to achieving complex optical modulation. With available technology, true complex modulation requires combining the action of multiple devices. For example, Optalysys have experience constructing systems which image a first SLM onto a second SLM, resulting in an optical action which is the product of that of the two SLMs. An illustration of such a system is shown.

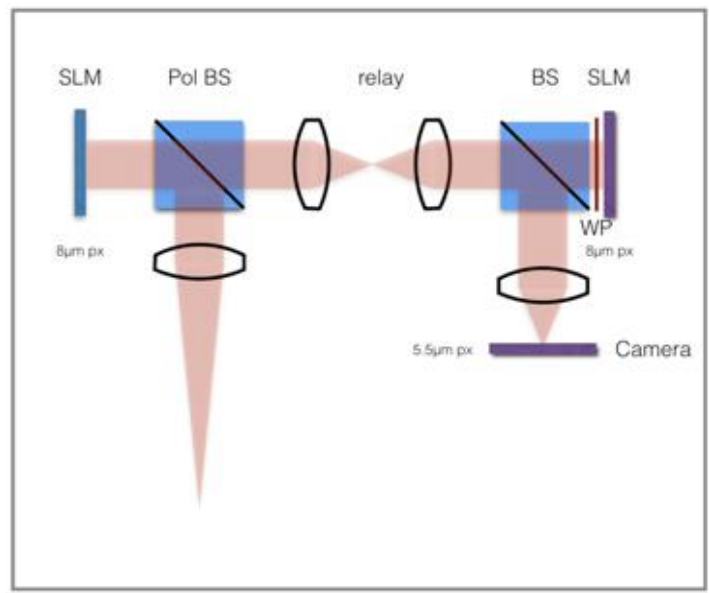

*A system which achieves complex modulation by combining the action of two SLMs. The polarisation scheme in this system is set so that the first SLM modulates predominately amplitude, and the second modulates phase.*

A key requirement of such a system is tight control of the polarisation, so that the operation of each SLM is well constrained. We have successfully constructed such systems – they were required by the previous work on performing the BiFFT properly – and are undergoing significant development.

A first hurdle is developing a high-performance optical relay. If the first SLM is not properly imaged onto the second, crosstalk occurs between the different pixels. The effect of light from a given pixel on the first SLM inadvertently landing on unintended pixels on the second SLM leads to spurious results. As pixels sizes for high-performance devices decrease, a very significant demand is placed on the optical relay. Fundamentally, the performance of this relay is limited by the optical diffraction limit.

A second hurdle is integrating multiple two-SLM systems together. For example, if both the input and filter are to offer arbitrary complex modulation, a system with 4 serial SLMs is required, adding extra optical surfaces to a system, which increases complexity and decreases fidelity.

### 4.4.13 Accommodating the sampling restrictions imposed by using an SLM

A further key restriction imposed by this optical method is the fact that the sampling function is imposed by the optical hardware. In particular, all available SLMs consist of a discrete rectilinear array of pixels. This is at odds with some of the continuous sampling functions used in NWP to represent weather quantities on the surface of the Earth.

The fundamental issue is that we wish to optically perform mathematical operations on data which is sampled over some non-rectilinear grid, but using optical hardware with regularly spaced rectilinearly arranged pixels.

An obvious solution to this problem is simply to interpolate from our data-sampling regime to a sampling regime appropriate for the optical hardware. While this is viable, it has attendant issues of compromising accuracy and an additional computational overhead.

We have developed a method which allows us to transfer the data directly from an arbitrary sampling grid onto a rectilinear SLM – with no modifications applied – and extract the desired numerical projection by applying modifications instead to the filter. Unfortunately, IP issues preclude disclosure of the specifics of this method on a public document at this stage.

However, as evidence of the utility of this method, we present simulation results showing the agreement between the 'correct' calculation of a projection onto a basis function, and the case where the data has been directly shown on a regularly spaced grid and the quantity extracted by an appropriately modified filter function.

*A comparison of extraction of numerical projections (a Fourier coefficient) from data on an irregular sampling grid (x-axis), and the same data placed directly on a regularly sampled grid (y-axis) with an appropriately compensated filter.*

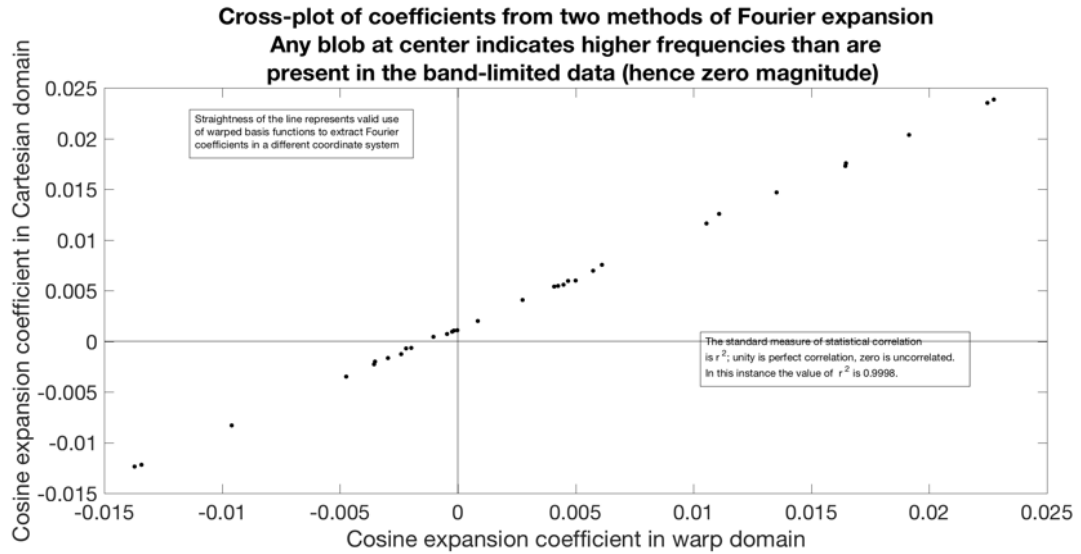

There are limits to the sampling discrepancies which this method can effectively compensate for, beyond those that would appear in any method. In the case of highly diverging pairs of grids, the filter cannot effectively compensate for the discrepancy.

Nonetheless, this technique is critical to broaden the utility of optical numerical processing techniques while still making use of the regular grids inherent in SLMs.

### 4.4.14 Discussion

#### 4.4.14.1 Implication of a spectral transform coprocessor on system performance

It is important to assess the utility of a hardware-specific spectral transform coprocessor in the context of a larger computational model, such as the integrated forecasting system (IFS) used by ECMWF. The system being proposed here is not a general-purpose computing module, but a very specific hardware offering.

Specifically, it is designed to increase the performance of the spectral transform component of the model.

Indeed, the technology offers the possibility of evaluating spectral transforms at very high resolutions exceptionally quickly. However, this would have significant implications for the model. To illustrate this, we will reproduce some profiling diagnostics of the MPI (Message Passing Interface) aspect of the IFS, acquired by the Barcelona Supercomputing Centre under Dr. Mario Acosta.

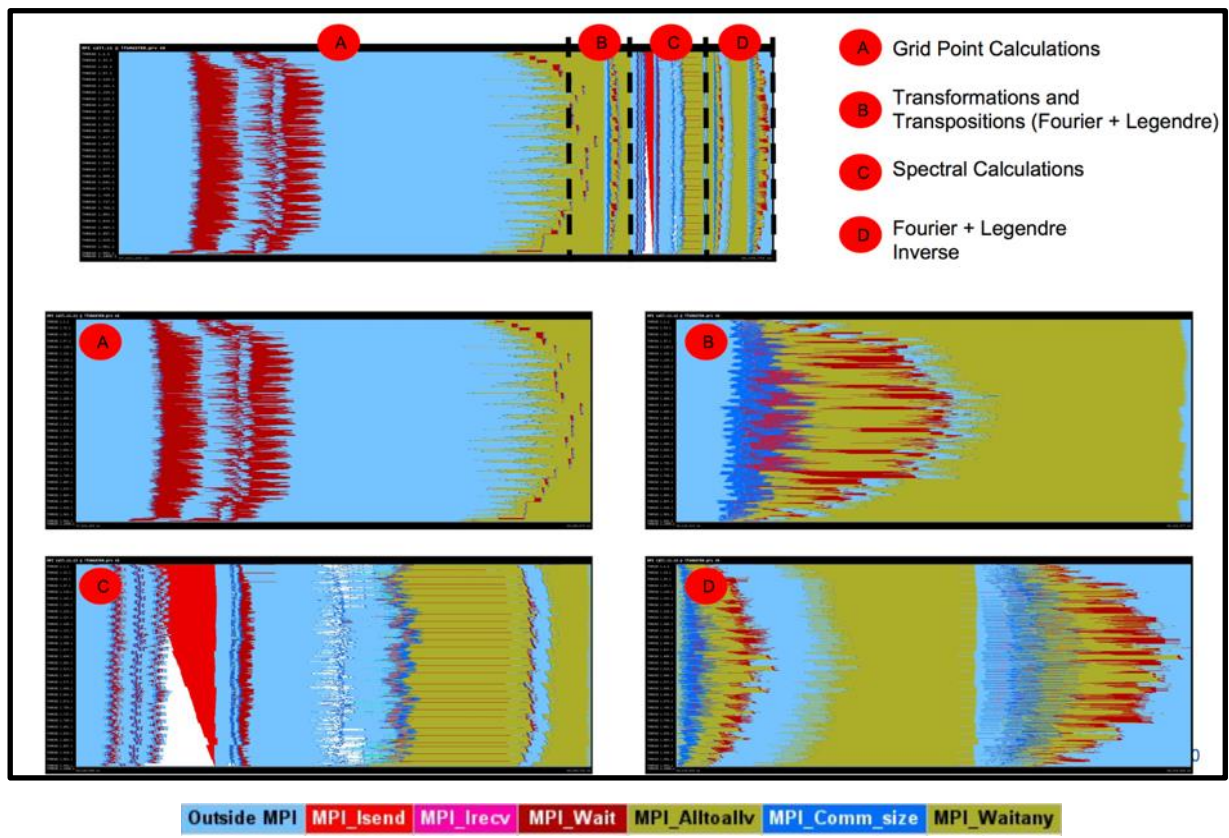

*A diagnostic of a single time-step of the IFS model, by capturing the different MPI processes. The horizontal axis is wall-clock time through the time-step; the vertical axis shows different processors in system.*

*(From "Profiling and Computational Performance of IFS using BSC Tools" by M. C. Acosta, 2nd ESCAPE Project Dissemination Workshop, Poznan, September 2017)*

Of particular note are the forward and reverse transform steps (B and D in the figure). It is clear that the *MPI_Alltoallv* process – which represents communicating the global state of the model between all of the different nodes – represents a very significant component of the spectral transform section. It is not the evaluation of the spectral transform itself which constitutes the main performance hurdle for global spectral models, but the communication bandwidth.

Hence, a high-performing hardware-specific spectral-transform coprocessor does not necessarily offer a significant advantage in terms of increasing model performance. Indeed, by facilitating – or arguably imposing – higher resolution transforms, the bandwidth burden is increased rather than decreased.

### 4.4.15 A use for a spectral transform coprocessor

As has been demonstrated, the use of a spectral transform coprocessor does not lend itself well to tasks requiring high numerical accuracy. However, this is not to say that a spectral transform coprocessor does not offer utility in the context of NWP. While the global communication burden in an active model causes issues, the post-processing of model data (converting spectral quantities back to grid-point quantities) could be facilitated by the increased raw performance of a spectral transform coprocessor.

This post-processing application is not a small one, and is key to making the model data useful to forecast users.

### 4.4.16 Applications of an optical correlator within NWP

The work under the ESCAPE program has been only part of the R&D conducted by Optalysys in this period. As a start-up looking to commercialize coherent optical information processing technology, we have been working across a number of different strategic developmental areas and there has naturally been cross-pollination between them.

The most fruitful area of development has been in developing a traditional optical correlator system, focused on a bioinformatics application. Specifically, the problem of genomic alignment – essentially noisy string matching of a relatively short string against a much larger genomic database. In order to support this application, we have had to develop a custom hardware drive solution for the SLMs and camera. This hardware possesses a significant latent performance with significant gains in power efficiency relative to traditional computing architectures. This capability will be displayed within the scope of the ESCAPE project under Work Package 4.

However, it is worthwhile in the present context to discuss this capability and outline ways in which this technology could potentially be of interest to the NWP community, albeit outside of the scope of the ESCAPE project itself. The most promising current avenue of investigation for this technology is in deep learning. Specifically, the optical correlator provides a potent platform on which to evaluate 2D convolutions. These operations are at the core of ConvNets (convolutional neural nets), which are the pre-eminent platform for image-based machine learning. Moreover, they represent the bulk of the computational burden when evaluating these neural nets. Furthermore, inference on neural nets is an application which is naturally tolerant of reduced precision computation, as offered by these optical techniques. By optically evaluating the convolutional layers of a deep-learning neural net, there is the possibility of offering a significant performance increase.

Neural nets have potential utility in NWP, either in the parameterisation of some physical effects, or data processing and inference.

### 4.4.17 Conclusion

Over the course of this project, we have advanced the capability of numerical optical processing, with particular reference to numerical weather prediction (NWP). In particular, we have developed optical techniques to implement a spectral transform which is not trivial optically: the spherical harmonics transform. This represents a significant innovation in the field of optical information processing. Two methods to evaluate this transform have been proposed, with increasing performance. The first makes use of an optical correlator architecture, demonstrating further the utility of this classic optical processing system. The second develops a novel multichannel astigmatic optical processor to combine two different optical stages.

There are significant engineering hurdles to deploying these systems with sufficient processing power and precision to be competitive with traditional methods. Nonetheless, we have identified and advanced the techniques required to perform these high precision operations optically.

Optical processing is more appropriately applied to cases where high-throughput relatively-complex operations are the priority, with less of an emphasis on numerical precision. The inherent ability of optical correlators to rapidly process convolutions naturally leads to the formation of convolution neural nets and machine learning technologies and holds the potential for aiding in the application of post-processing NWP data.

## 5 Dwarf 2: MPDATA

### 5.1 GPU Optimizations

#### 5.1.1 Introduction

In this section we present our work to optimize execution of the most computationally expensive kernel in the MPDATA dwarf, `compute_fluxzdiv`. We use the "dwarf-D-advection-MPDATA-solidBody-pole-O128" test case, chosen to be small enough to fit on a single GPU but large enough to give kernels of a realistic size when running on that GPU. We use the NVIDIA Tesla P100 GPU for all results. We first describe the original baseline GPU implementation of the kernel, go on to describe the steps taken to optimize, and finally present performance results.

#### 5.1.2 Baseline performance

The most computationally expensive kernel in MPDATA was identified as `compute_fluxzdiv`, through profiling on the CPU where it is responsible for 25% of overall runtime. This kernel has the following structure:

```
1.  do jnode  = 1,this%geom%nb_nodes
2.   do jlev = 1,this%geom%nb_levels
3.     zsum = 0.0_wp
4.     do jedge = 1,inode2edges_size(jnode)
5.       iedge = inode2edges(jnode, jedge)
6.       zadd = real(this%geom%node2edge_sign(jedge,jnode),wp)
7.       zsum = zsum+zadd*pFx(jlev,iedge)
8.     enddo
9.     pdivVD(jlev,jnode) = zsum/pvol(jnode)              &
10.      & +(pFz(jlev+1,jnode)-pFz(jlev,jnode))/this%dz
11.  enddo
12. enddo
```

It can clearly be seen that there is a relatively low arithmetic intensity (where for more details on this please see the Spherical Harmonics GPU section), so global memory bandwidth will be the limiting factor for test cases of realistic size and the performance "roofline" can be taken as the memory bandwidth achieved by the STREAM benchmark.

The initial GPU port involved the enclosure of the entire loop nest in OpenACC `kernels` start and end directives, which instructs the compiler to parallelise the whole region in the way it best sees fit. Additionally, there existed an OpenACC `loop` directive before the loop at line 4 with a reduction clause on the `zsum` variable, to instruct the compiler to ensure that the necessary reduction is performed. This resulted in the compiler deciding to assign the outermost loop (at L1) to CUDA blocks, the loop at L2 to CUDA threads within each block and the innermost loop (at L4) as a sequential loop performed by each thread.

The resulting performance of this implementation was low: it is only able to achieve 44GB/s data throughput on the NVIDIA Tesla P100 GPU, which is less than 10% of that achieved by the STREAM benchmark. The reasons for this are related to suboptimal parallel decomposition and data layout, and the accessing of data in deep structures: the following section will describe optimizations which overcome these limitations.

### 5.1.3  Optimization strategy

As can be seen the above code snippet, there exist three nested loops which must be mapped to the parallelism of the GPU. The extents of these, for the test case in use, are the following

- L1 (nb_nodes): 71424
- L2 (nb_levels): 3
- L4 (nb_edges): 213199

There are a number of possibilities for how the parallel mapping can be implemented and the original choice was suboptimal. The reason for this is that the loop assigned to CUDA threads within each block (at L2) has an extremely small extent, where typically we need much more parallelism at the CUDA thread level to make good use of the vector nature of the CUDA execution model. Instead, we choose to collapse the two outermost loops (through use of the OpenACC `parallel loop collapse(2)` directive) and assign the parallelism across this collapsed loop to both CUDA blocks and threads within each block. This allows the compiler to decide a much more suitable extent of vectorization. An OpenACC `loop seq` directive is applied to the innermost loop, such that each thread will perform all of this loop in a sequential manner (satisfying the requirements of the reduction). With this new parallelisation strategy, it is important to ensure that data is accessed in a coalesced manner for the field arrays `pFx, pFz` and `pdivVD`, in order to achieve a high percentage of memory bandwidth. For coalescing, we need consecutive threads (corresponding to consecutive `jlev` indices) to access consecutive memory addresses, and fortunately the original data layout of these arrays (with `jlev` the fastest moving innermost index in Fortran) already satisfies this requirement, so no further data layout modifications are necessary.

The kernel accesses several read-only data elements and structures. For these, best performance is achieved when the compiler maps the data to the fast on-chip constant cache on the GPU. However, we find that, for the case of the deep array access `this%geom%node2edge_sign` the compiler does not make full use of this capability. But, if we copy this to a regular "flat" array, ahead of kernel execution and use this in place of the original structure we see an increase in constant cache utilization and improved performance. Furthermore, we can see that the operation involves division by a constant (`this%dz`). We replace this by multiplication by the reciprocal of the constant (calculated in advance), which futher boosts performance.

### 5.1.4 Performance results

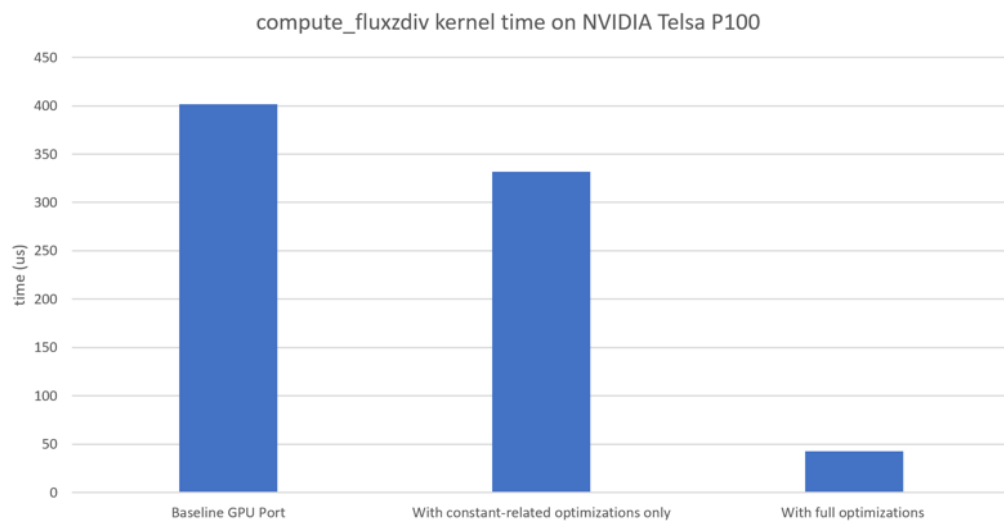

*Figure 16: The time taken by the original and optimized GPU versions of the compute_fluxzdiv kernel on the NVIDIA Tesla P100 GPU.*

In Figure 16 we see the effects of the optimizations described in the previous section, which have decreased the time taken by the kernel by a factor of 9.4x. The achieved throughput of the optimized version is measured by the NVIDIA profiler to be 344GB/s, which is 66% of the value measured using STREAM benchmark, indicating that we are reasonably close to the hardware limit (but there may be scope for some further optimization). The figure also includes an intermediate result to show the effect of the optimizations related to constant data (i.e. the replacement of the deep array and replacement of the division by the multiplication of a reciprocal): it can be seen that these are significant but modest in comparison to the more important parallel decomposition tuning.

### 5.1.5 Summary

In this section, we presented our work to optimize the most computationally demanding kernel in the MPDATA dwarf: `compute_fluxzdiv`. The kernel is memory-bandwidth bound, and the original GPU implementation showed poor performance at less than 10% of achievable bandwidth (as measured by the STREAM benchmark). The optimizations involved improving the parallel decomposition on the GPU, the replacement of a deep array containing constant data with a flat alternative and the replacement of division by multiplication using the reciprocal of the original constant. The time taken by the optimized kernel is 9.4x lower than the original baseline GPU version and is now performing much closer to the roofline at 66% of achievable bandwidth.

## 5.1 CPU Optimizations

*CPU and Xeon Phi optimizations of MPDATA have been reported in Deliverable D3.2. For details on these results see the corresponding report.*

## 6 Dwarf 3: Radiation

### 6.1 Introduction

This chapter is a summary of the results in the paper "Tuning the implementation of the radiation scheme ACRANEB2" which was published in the 9th edition of the combined Newsletter of the HIRLAM and ALADIN consortia. The paper is available here: http://www.dmi.dk/fileadmin/user_upload/Rapporter/TR/2017/SR17-22.pdf.

In this part of the project, we did not only strive to achieve optimal performance for a dwarf on a given architecture, but also attempted the find a representation that achieved high performance with the same code base.

We found that one can indeed refactor the code such that it runs with competitive performance on modern throughput architectures such as NVIDIA GPUs and Intel Xeon Phis. The parallelism itself is expressed using directive based approaches, OpenMP and OpenACC, respectively. We show that competitive performance is easiest to achieve with different code bases. Portable performance, i.e. using same data structures and same overall code base, is attainable too but at this moment and with the current level of maturity of the toolchain the resulting performance may not be competitive.

The dimensions of the primary testcase that was used for this dwarf were 400x400x80. The provided reference timing of the baseline code run on the 400x400x80 test case was more than 1600 seconds on a SNB node (E5-2680v1) but a simple performance projections suggested that it should be in the order of a few seconds on modern hardware.

## 6.2 Baseline performance, GPU

Establishing a baseline timing of the radiation kernel on the GPU was not straight-forward. Clearly, the baseline code was not suited for the modern throughput architectures. Thus, just to get the baseline code on an NVIDIA GPU, we had to make some non-trivial code changes (amounting to a 200+ lines of code in the most expensive routine alone) to ensure the semantics ended up being correctly understood by the compiler. In addition, we had to use a smaller testcase to fit into the GPUs' memory.

The resulting baseline code, running only a single thread on the entire GPU, runs ~4500x slower on an NVIDIA P100 from 2016 than the pure baseline code on a SNB from 2012. Obviously, the code needs to be instrumented with parallelization directives to properly utilize a GPU.

## 6.3 Baseline performance, CPU

On Intel Xeon and Xeon Phi the baseline code ran correctly out of the box. However, the baseline code was not threaded.

For single core, the baseline code on the 2016 Xeon technology (BDW) beats the 2012 Xeon technology (SNB) only by a factor of 1.7. With the baseline code, KNL-7210 from 2016 runs ~2 times slower than a SNB from 2012.

### 6.3.1 Optimization strategy

Clearly, one cannot take a legacy code developed for older CPUs or vector machines, decorate it with directives and expect high performance gain on modern CPUs or GPUs. We do not regard this as a flaw of the directive based programming models but rather see it as a question of legacy. The same legacy issues would arise had we chosen another programming model such as e.g. CUDA and OpenCL.

The paper takes two different approaches to the code tuning: One where the overall data structures are retained and another where we focus solely on the strengths of a particular architecture. This result in three different code bases:

1. X: portable data-structures, tuned for Intel Xeon/Xeon Phi

2. G: portable data-structures, tuned for GPU

3. GNM: non-portable data-structures, tuned solely for the GPU

This is all described in more details in the paper.

### 6.3.2 Performance Results GPU

P100 timings of the G and GNM codes on the 400x400x80 test case were 4s and 1.7s, respectively, which perfectly matches our naive expectations from the introduction. The speedup achieved on P100 by refactoring the code is ~7300x for the G code and

~17000x for the GNM code. This huge speedup should be seen in the light of inherited legacy.

### 6.3.3 Performance results, CPU

On a single core, our refactored version runs 3.3x and 3.8xfaster, respectively, than the baseline code on a SNB from 2012 and on a BDW from 2016.

In fully threaded context on SNB (E5-2680v1) the refactored code runs 50 faster than the baseline code, while the corresponding speedup is 110 on the BDW (E5-2699v4). These speedups are more than we would expect from the threading alone (i.e. ideally 32x and 88x, respectively).

Single core performance of the refactored code is more than 6 times better on KNL than on SNB. The refactored code runs 11x faster than the baseline code on a single KNL core. This is attributed to a better utilization of the ISA in the refactored code.

The effect of refactoring is a factor of 667x on KNL-7210 which again is more than the threading to 256 threads alone would give us.

### 6.3.4 Summary

The completely refactored codes give correct results on all the tested hardware and with all compilers tested across all incarnations (testcase size, thread count, etc).

NODE PERFORMANCE IMPROVEMENT FACTORS RELATIVE TO SNB. FOR P100, (A) AND (B) ARE WITHOUT AND WITH ALGEBRAIC REWRITE OF THE POWER FUNCTION, RESPECTIVELY.

| Architecture | HPL | Stream Triad | transt3 |
|---|---|---|---|
| E5-2680v1 | 1.0 | 1.0 | 1.0 |
| E5-2699v4 | 4.2 | 1.6 | 4.5 |
| KNL-7210 | 5.6 | 5.6 | 6.7 |
| KNL-7250 | 5.7 | 6.2 | 7.8 |
| NVIDIA-P100 (a) | 11.4 | 6.9 | 7.6 |
| NVIDIA-P100 (b) | 11.4 | 6.9 | 9.5 |

*Table 3 Performance improvements relative to SandyBridge, incl. Stream Performance*

In comparing with the baseline code, we have seen vast speedups on all platforms primarily as a result of the legacy state of the code base. These speedups are, not surprisingly, most profound on the new bandwidth optimized platforms. Our refactorization efforts have led to a code bases that exceeds the performance expectations dictated by Moores Law, i.e. we exceed the performance factors dictated by both stream and HPL when running on hardware that emerged in the period 2012-2016. This is shown in Table 3.

REFACTORIZATION IMPROVEMENT FACTOR ON A SINGLE NODE AND ON
A SINGLE CORE FOR DIFFERENT ARCHITECTURES.

| Architecture | Core | Node |
|---|---|---|
| E5-2680v1 | 3.3 | 50 |
| E5-2699v4 | 3.7 | 110 |
| KNL-7210 | 11.0 | 667 |
| NVIDIA P100 | N/A | 7302 |
| *NVIDIA P100* | N/A | *17000* |

*Table 4 Performance improvements by individual node refactoring.*

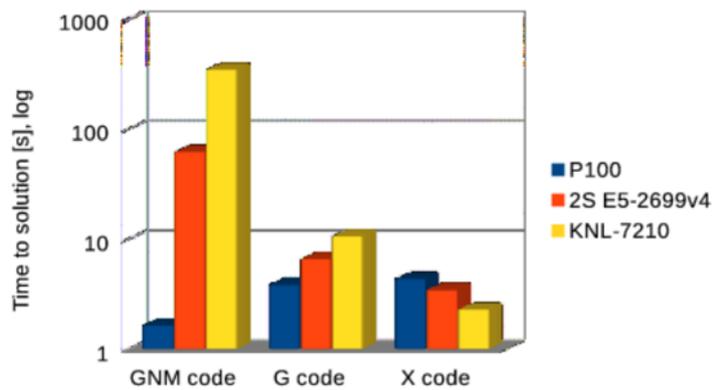

*Figure 17 Impact of different data layouts on overall performance. It turns out that the optimal performance on GPU requires a different data layout than on a CPU.*

Refactoring of legacy code is increasingly more important as shown in Table 4 and Figure 17. Achieving best performance on different architectures currently requires different code bases for this radiation scheme.

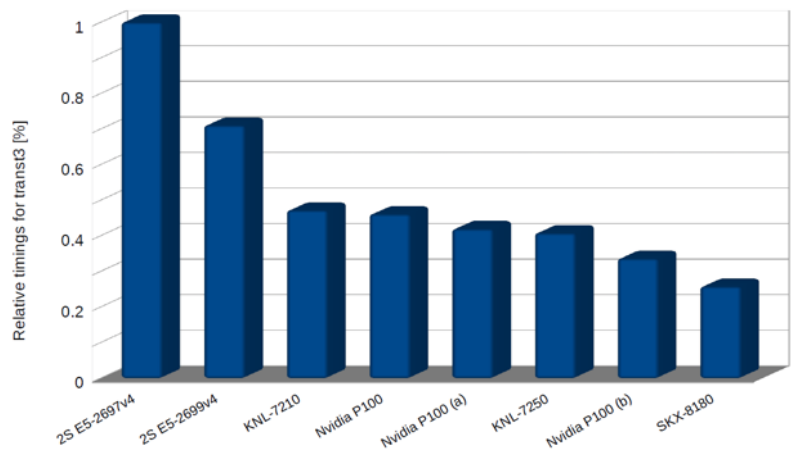

*Figure 18 Relative timings for the different implementations of the radiation scheme.*

Figure 18 summarizes the best performance on architectures from 2016 plus an out-of-the-box timing of the X code on a Skylake from 2017. With the P100 (b) timing of 1.7s, the initial simple performance projections of a few seconds on modern architectures were met with our refactored code bases.

# 7 Dwarf 5: Spectral Transform – Bi Fourier

## 7.1 CPU Optimizations

### 7.1.1 Baseline performance

The Spectral Transform Bi Fourier dwarf has been profiled using a variety of profiling tools (Intel MPS, Intel Advisor, Intel Vtune, Allinea MAP and DDT) on both Intel Xeon and Intel Xeon Phi 7250 (KNL).

#### 7.1.1.1 Profiling on Xeon

This sub-section presents the results of an early profiling on a system based on Xeon processor E5-2650V3 (see Table 1 for its main characteristics). A detailed study is available in a dedicated presentation available on the ESCAPE Confluence site. This work has been done on single node using 10 MPI tasks and 2 OpenMP threads per MPI tasks (HT has not been used; we used all physical cores). The test case used is data_bifft_200x180.grb with 1000 iterations (ITERS). Finally the dwarf has been compiled in order to use AVX2 ISA.

As shown by Figure 19 from MPS tool the OpenMP parallelization is not good, as the serial time represents more than 66% of the execution time in addition to a high OpenMP imbalance. Moreover it clearly appeared that the dwarf is memory bound which lower vectorization.

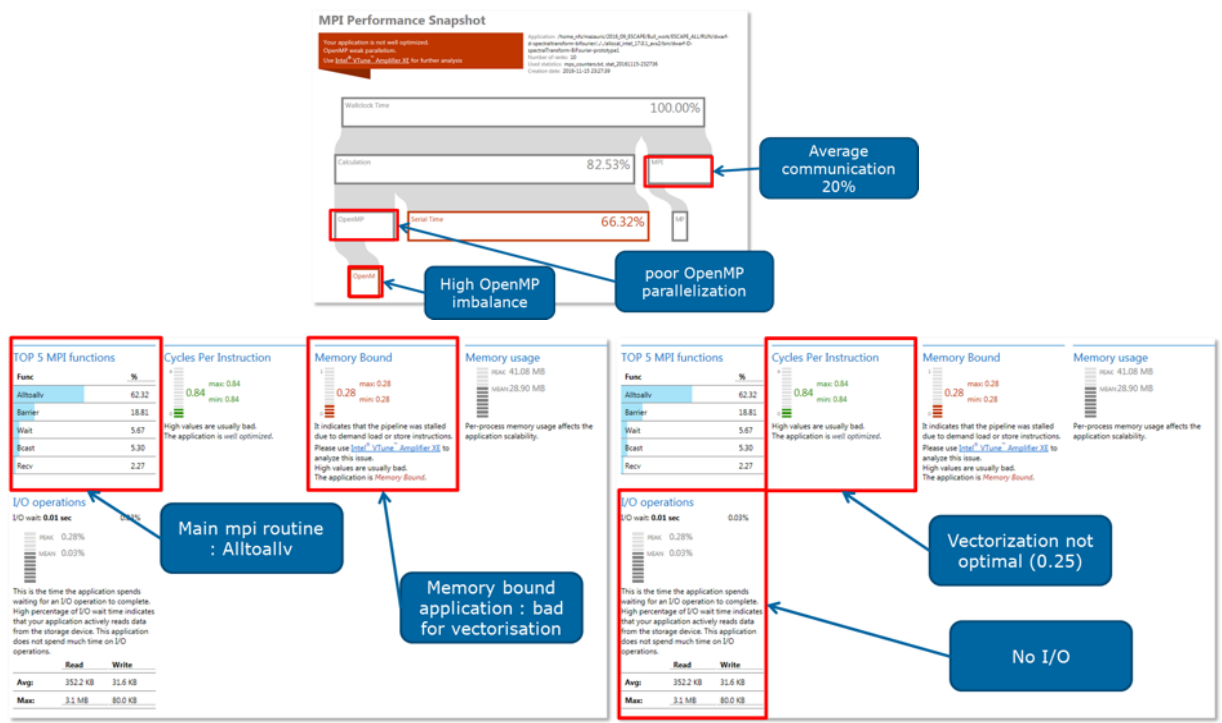

*Figure 19 – Spectral Transform – Bi Fourier profiling result (MPS screenshots)*

Vtune profiling (snapshot given Figure 20) reenforced this analysis, as most of the time is spent in the Intel OpenMP library (especially in the barrier function). Next, the top time-consuming functions are named *qpassf* and *rpassf*.

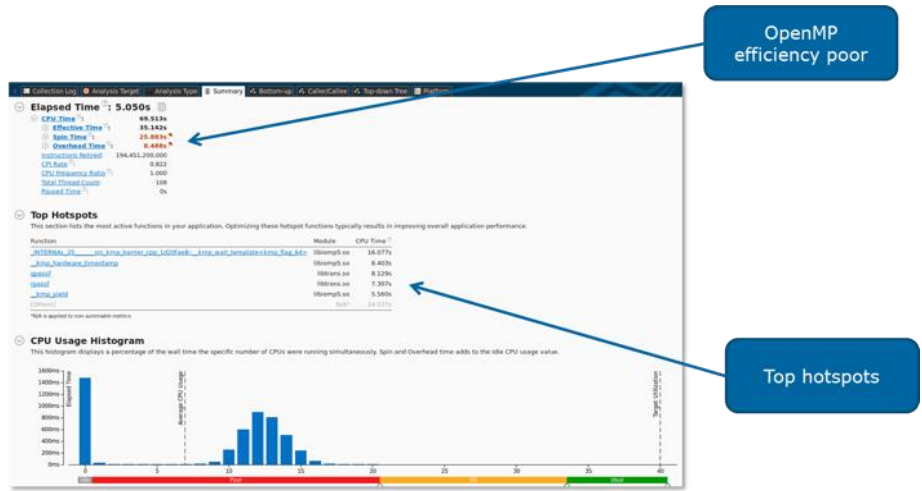

*Figure 20 – Spectral Transform – Bi Fourier profiling result (Vtune screenshot)*

Concerning the vectorization, a low part of the dwarf is vectorized (only 10 loops leading to more than 95% of the time spent in scalar mode).

### 7.1.1.2  Profiling on Xeon Phi

This sub-section presents the results of an early profiling on a system based on Xeon Phi processor (see KNL 7250 Table 5). The dwarf has been compiled in order to use AVX-512 ISA.

| Processors type | KNL 7250 |
|---|---|
| Frequency | 1,4 GHz |
| #cores | 68 |
| #threads per core | 4 |
| KNL mode | Quadrant cache |
| Memory | 6x16 GB + 16 GB (MCDRAM) |

*Table 5 – Xeon Phi (KNL) 7250 main characteristics*

The roofline graph obtained with Intel Advisor shows that the dwarf doesn't exploit the memory bandwidth available. This means that deep modifications are needed to improve data locality through for example restructuring the main loop and the data structure or adding cache blocking.

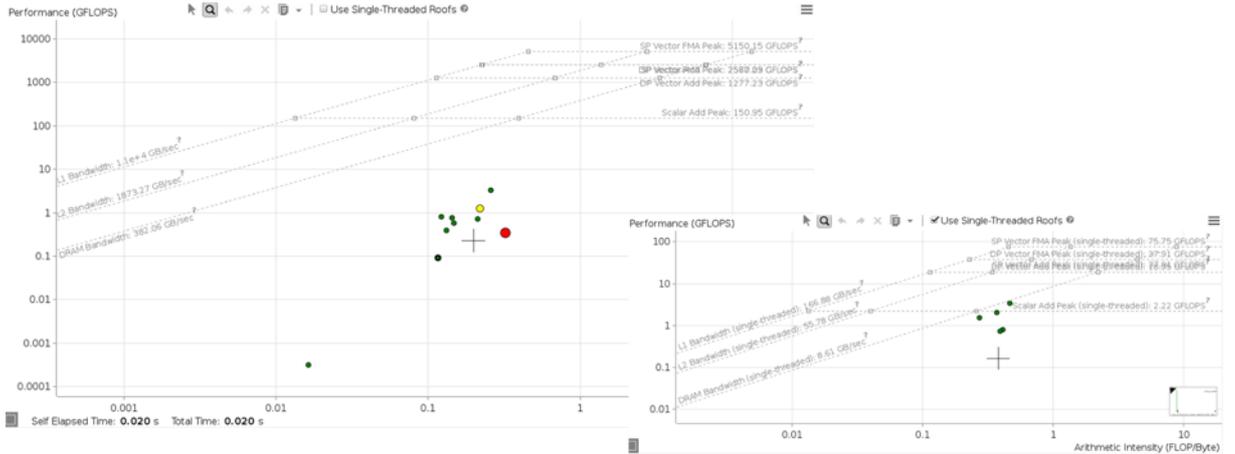

*Figure 21 – Spectral Transform – Bi Fourier roofline graph from Intel Advisor on KNL 7250. The left graph represents all treads roofs and the left graph represents single thread roofs.*

Concerning the vectorization of this dwarf, Intel Advisor showed that only 9% of the loops are vectorized and the vectorization efficiency is about 33%. It also advised that *qpassf* and *rpassf* should be vectorized (see Figure 22).

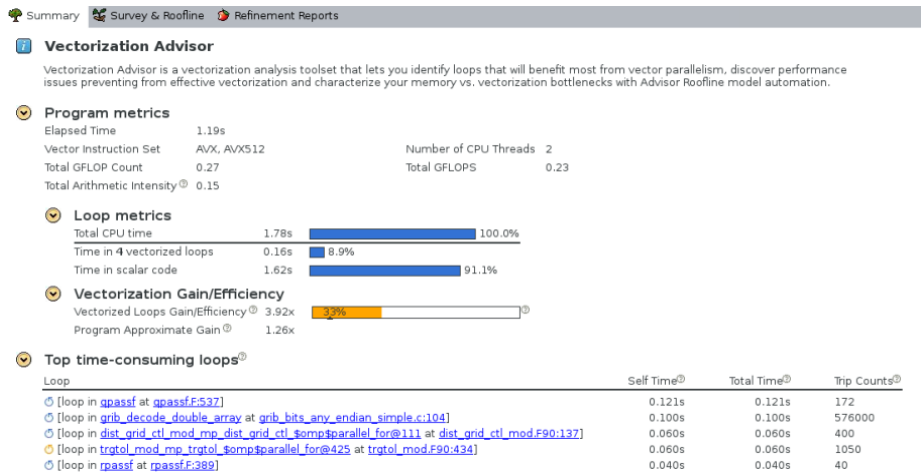

*Figure 22 – Vectorization analysis of Bi Fourier dwarf.*

A finer profiling of the dwarf is presented in the following Figure 23, highlighting 5 main hotspots representing 40% of the time. Complex and non-uniform memory access via indirections or data dependencies, intra and inter iteration, avoid compiler parallelisation and vectorization.

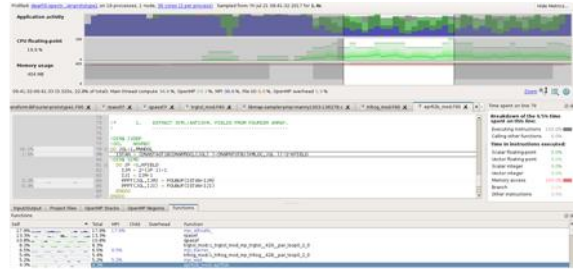

- ► 5 main hotspots : 40% of time

- ► trans/src/cpu/algor/fourier/rpassf.F
- ► trans/src/cpu/algor/fourier/qpassf.F

- ► trans/src/cpu/trans/module/trgtol_mod.F90
- ► trans/src/cpu/trans/module/trltog_mod.F90

- ► etrans/src/etrans/module/eprfi2b_mod.F90

```
DO 430 IJK=1,ILOT
    C(JA+J)=(A(IA+I)+A(IC+I))+(A(IB+I)+A(ID+I))
    D(JA+J)=(B(IA+I)-B(IC+I))+(B(IB+I)-B(ID+I))
```

```
DO JK=IFIRST,ILAST
    IPOS = INDEX( INDOFF_val+JK )
    PGLAT(JFLD,IPOS) = PGP(JK,JFLD,JBLK)
```

```
ISTAN = (D%NSTAGT1B(D%NPROCL(JGL) )+D%NPNTGTB1(KMLOC,JGL))*2*KFIELD
```

*Figure 23 – Hotspot analysis of Bi Fourier dwarf.*

### 7.1.2   Optimization strategy

Optimization strategy focused on improving vectorization of the 5 identified hotspots by the use of compiler directives.

### 7.1.3   Performance results

On the Xeon, a gain of 17.7% in time has been obtained by adding SIMD directives on the parallel loops. According to Intel Advisor, the number of vectorized loops grew from 10 to 24. The time in scalar mode dropped from 95.5% to 80.6% but at the same time the vectorized loop gain/efficiency was reduced from 51% to 41%. This means that vectorization is not efficient due to memory stalls and the parallelism is not high enough.

The performance results are presented in the following graphs for Xeon Phi (Figure 24). Up to 16% of gain compared to the original performance has been achieved by adding SIMD directives. Same conclusion can be drawn for the Xeon but is restricted by the fact that AVX-512 registers are two times wider than AVX2 ones which gives poorer vectorization efficiency.

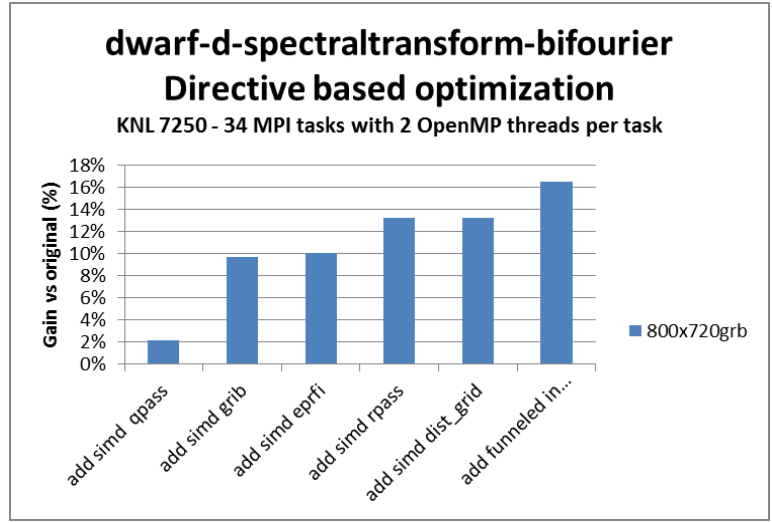

*Figure 24 - Performance result of the directive based optimization on the Bi Fourier dwarf*

### 7.1.4 Summary

In this section, a detailed profiling and benchmarking of the spectral transform – bi fourier dwarf has been presented, in addition to a set of non-intrusive optimization and their corresponding results. This original version of the code suffers from a low OpenMP efficiency and is memory bound. Improving the vectorization of the 5 identified hot spots led to a gain of 18% and 16% on the Xeon and Xeon Phi respectively. As a conclusion, improving the OpenMP efficiency and data locality represents the more relevant optimization tracks to achieve better performance. This will imply more intrusive optimization.

## 8 Dwarf 6: Cloud Microphysics IFS Scheme

### 8.1.1 Baseline performance

The intra node scalability on different processors and configurations for the cloud microphysics IFS scheme dwarf is presented in this section. As shown in Figure 25 the dwarf takes advantage to the hyper threading up to 2 threads by core on the SMP node (see E7-8890 v4 in Table 2, also called mesca2) with the lower execution time equal to 575 ms, followed by the E5-2690V4 processor with 590 ms. Among the different KNL configurations, the KNL2 one (cache mode/quadrant) is the best with 1106 ms.

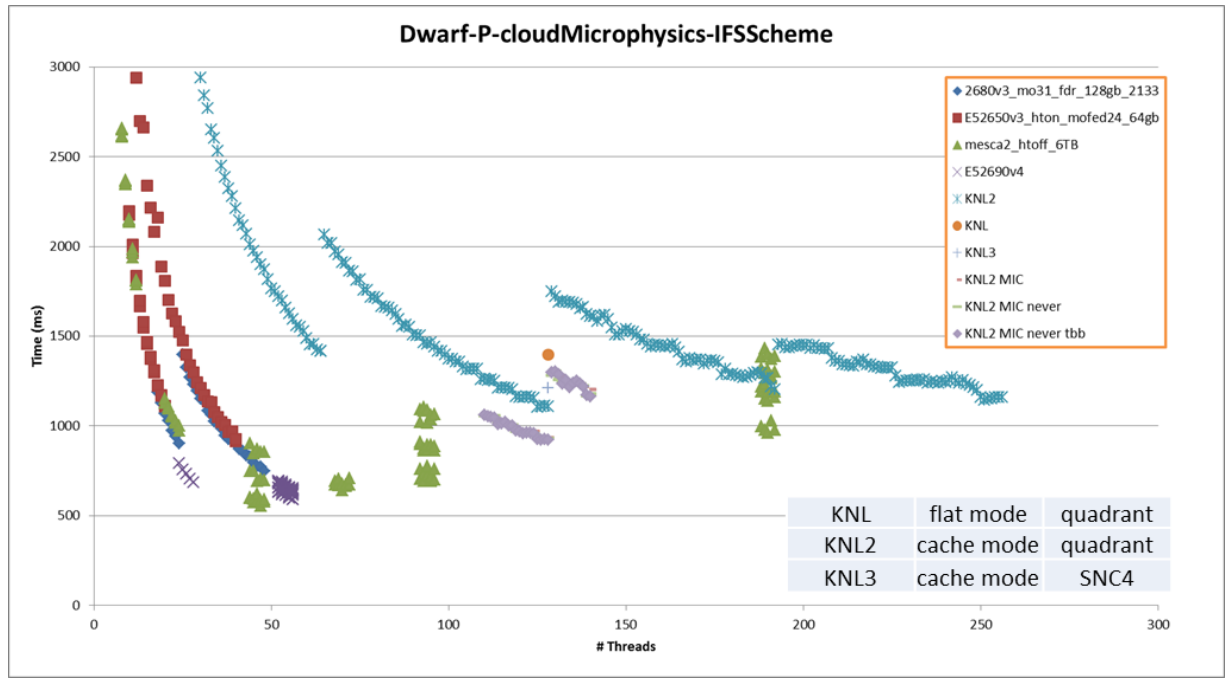

*Figure 25 - Cloud Microphysics IFS Scheme dwarf - Intra node scalability comparison on different system*

The roofline graph obtained with Intel Advisor (Figure 26) shows that the dwarf doesn't fully exploit the memory bandwidth available. This means that memory accesses should be optimized to improve data locality and vectorization efficiency and thus global performance.

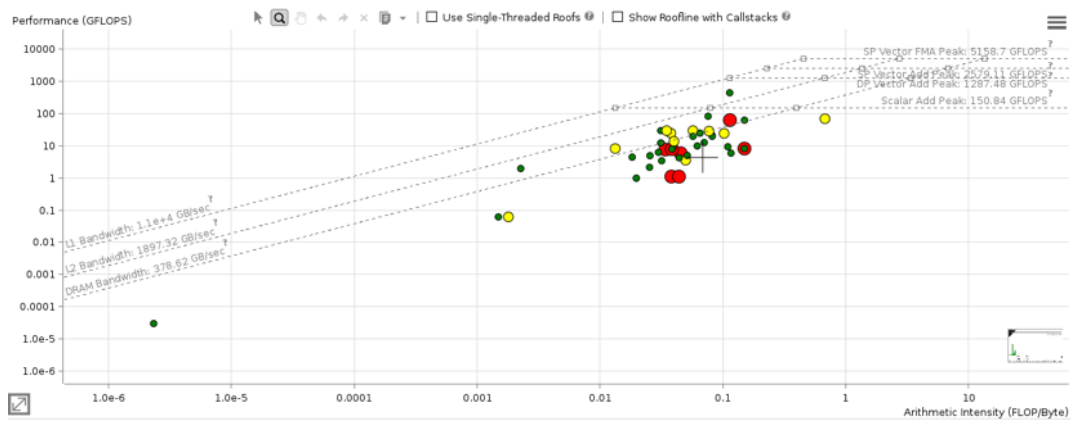

*Figure 26 – Cloud Microphysics IFS Scheme dwarf roofline graph from Intel Advisor on KNL 7250*

### 8.1.2  Optimization strategy

This dwarf contains a memory blocking scheme (parameter NPROMA) and the problem size can be parametrized (NGPTOT parameter). The first optimization was to find the best NPROMA according to NGPTOT and to the number of OpenMP threads. Then, according to the profiling results, memory optimisation through the impact of transparent huge page has been studied. In addition, directive based optimization to improve at the same time vectorization and memory alignment have been then applied, and finally minor code modification, to show how non-intrusive optimizations can bring performance improvements.

### 8.1.3 Performance results

#### 8.1.3.1 Parameter study

As shown Figure 27, the best parameters on a KNL 7250 are: NPROMA=64 with 134 OpenMP threads (128 threads give timings closed to 134). These figures also enforced the previous result: hyper threading (2 threads per core) improves performances.

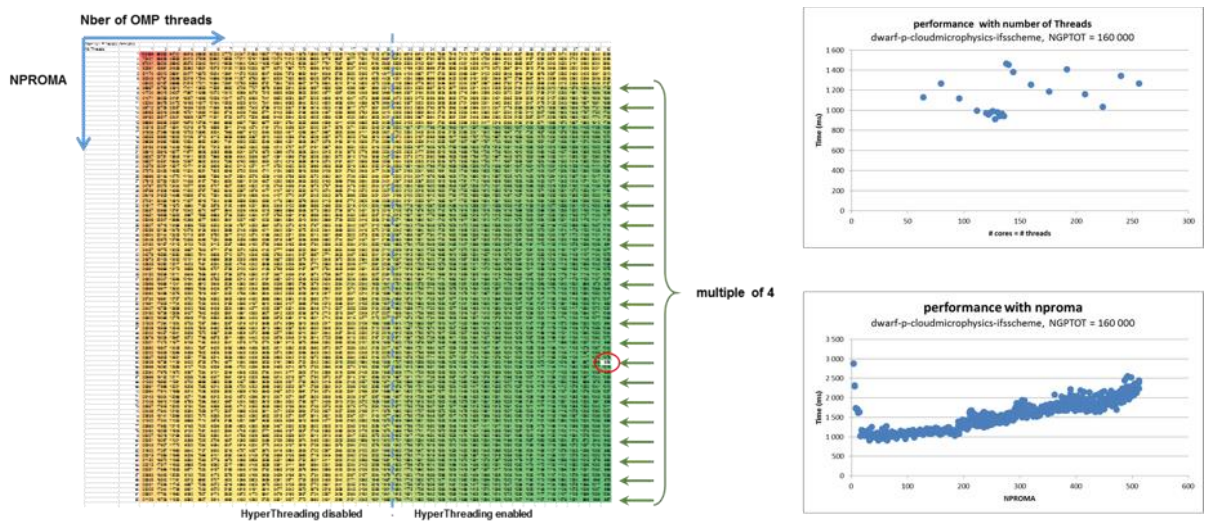

*Figure 27 – Cloud Microphysics IFS Scheme dwarf - Parameter study results. Left graph proves the interest of hyper threading (HT) and NPROMA should be multiple of 4 (on an E5-2650V3). Right graphs represent time results for the best parameters: NPROMA=64 with 128/134 threads on KNL.*

#### 8.1.3.2 Transparent Huge Page impact

This dwarf is very sensitive to transparent huge page as the execution time on a KNL 7250 dropped from 911 ms to 756 ms representing a gain of 17% (NGPTOT=160.000, 128 threads OpenMP, NPROMA=64).

#### 8.1.3.3 Memory alignment and vectorization with compiler directives

The first set of directive based optimization in addition with a good binding reduces the execution time from 756 ms to 693 ms representing a gain of 8.3% as shown Figure 28.

|  | # Nodes | # threads | Time (ms) |
|---|---|---|---|
| **Original CPU** | 1 | 128 | 756 |
| **add contiguous pointer in src** | 1 | 128 | 731 |
| **+ export KMP_HW_SUBSET=64c,2t** | 1 | 128 | 722 |
| **+ add simd pragma in cloudsc (L2884)** | 1 | 128 | 708 |
| **+ optim cloudsc** | 1 | 128 | 693 |

*Figure 28 - Cloud Microphysics IFS Scheme dwarf – Directive based optimization results (first set)*

In order to improve the previous optimizations, new compiler flags have been used giving a gain of 30% on the execution time. Alignment directive (other than contiguous) brings 7% more gain and optimization of the microphysics solver few more ms. The execution time of the most optimized version on KNL 7250 is 484 ms representing a speedup closed to 2x, better than the best original CPU execution time.

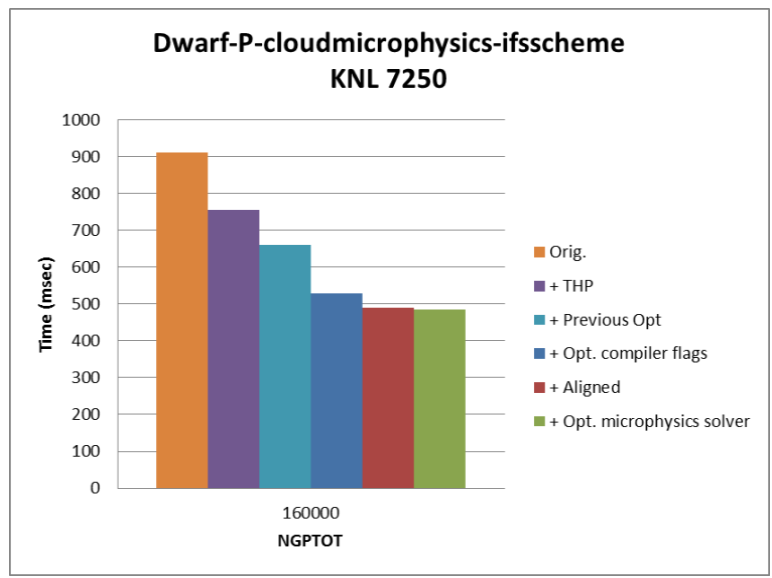

*Figure 29 - Cloud Microphysics IFS Scheme dwarf – Optimization results.*

### 8.1.4 Summary

In this chapter, a profiling and benchmarking of the cloud microphysics IFS scheme dwarf has been presented in addition with a set optimization (mainly non-intrusive) and their corresponding results. First an intra node scalability study on different processors and configurations has been presented. A parameter tuning, the use of transparent huge page, the incorporation of compiler directive and optimization flag tuning to improve compiler efficiency result in a global speedup closed to 2x on KNL.

At this point of the optimization work, a better understanding of the global algorithm seems to be required to envision future optimization tracks through algorithm and code modifications.

# 9   Dwarf 7: Elliptic solver GCR

## 9.1   CPU Optimizations

### 9.1.1   Baseline performance

Two versions of the Elliptic solver GCR dwarf have been studied: the first version of the code executed on Xeon (E5-2650V3) and a recent version on Xeon Phi (KNL 7250). This last version was not compilable using Intel compiler at the time of writing so (due to an internal compiler error still under investigation), so the GNU compiler toolchain has been used instead.

The results of the intra-node scalability study on Xeon are presented Figure 30. The maximum speedup is less than 2.5x meaning that the OpenMP parallelization is not efficient.

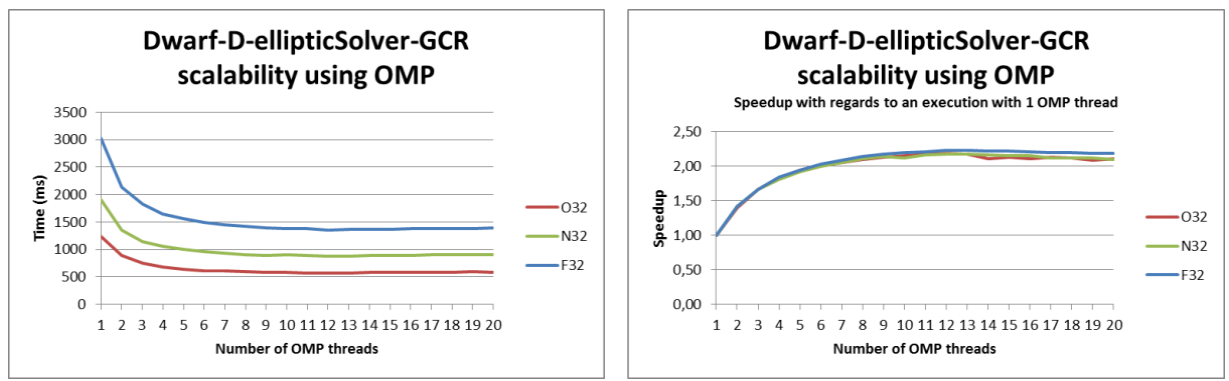

*Figure 30 - Elliptic solver GCR dwarf – OpenMP and iIntra node scalability results (Xeon).*

*Concerning the Xeon Phi, the results are presented*

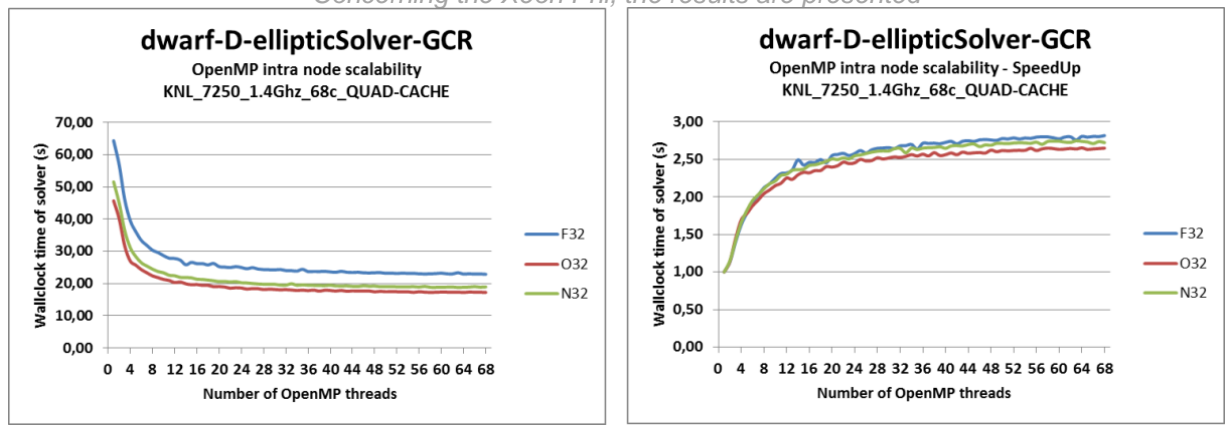

Figure 31. The maximum speedup is less than 3x meaning that the parallelization is also not efficient on KNL which was not a surprise. This time the MPI parallelization

on a single node has been benchmarked, but as one can see on Figure 32 it doesn't improve the performance, may be due to a misunderstanding on its usage.

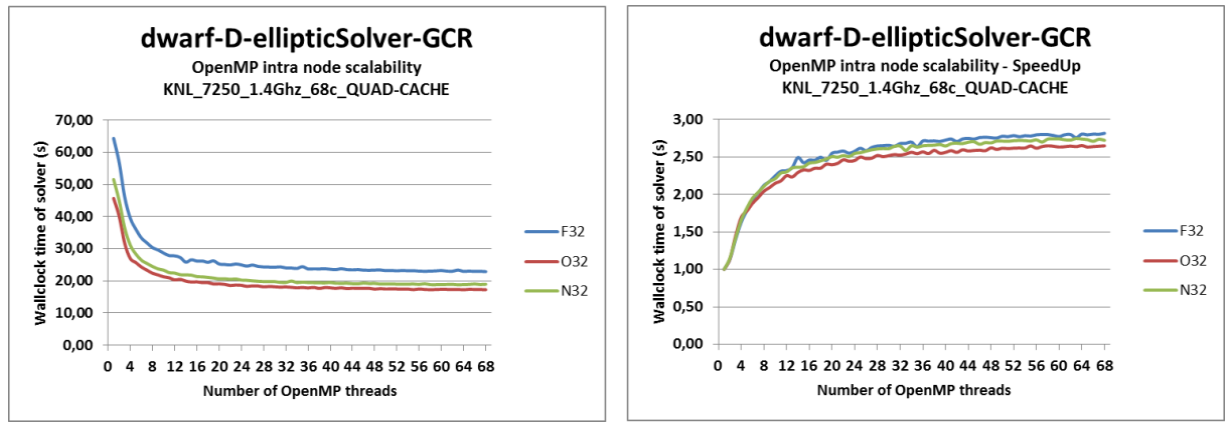

*Figure 31 - Elliptic solver GCR dwarf – OpenMP intra node scalability results, timing and speedup (Xeon Phi)*

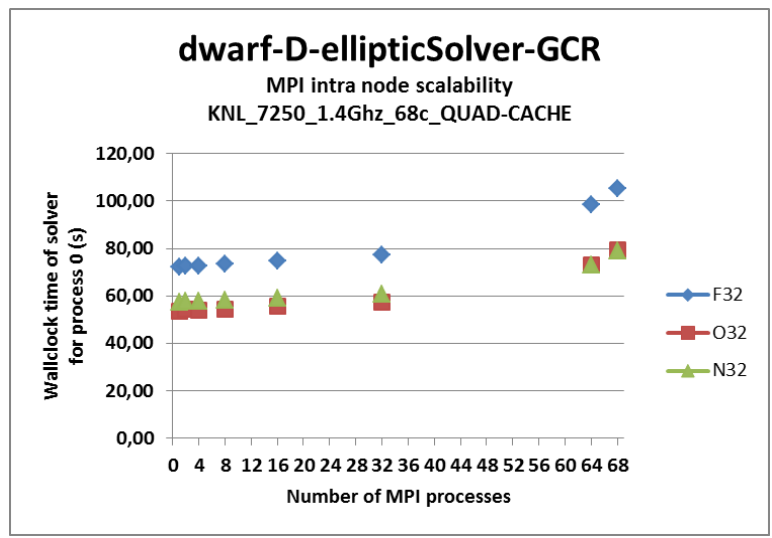

*Figure 32 - Elliptic solver GCR dwarf – MPI intra node scalability results (Xeon Phi)*

After a deeper profiling at both function and loop levels, it appeared that the most consuming execution time functions were "nabla operators divergence 2d" and "nabla operators divergence 3d" located in the "nabla operators" module.

### 9.1.2   Optimization strategy

The optimization strategy chosen focused on the optimization of the identified most time consuming functions. Two simple transformations of code hoisting has been done on two nested loops and a division has been also hoisted to enclosing loop and replaced by a multiplication in its original loop.

### 9.1.3  Performance results

The optimizations performance results in terms of execution time and speedup are presented Figure 33 for both mono and multi-threads OpenMP execution. The speedup obtained on the mono thread executions is up to 1.5x and up to 2.5x on the multi-thread ones.

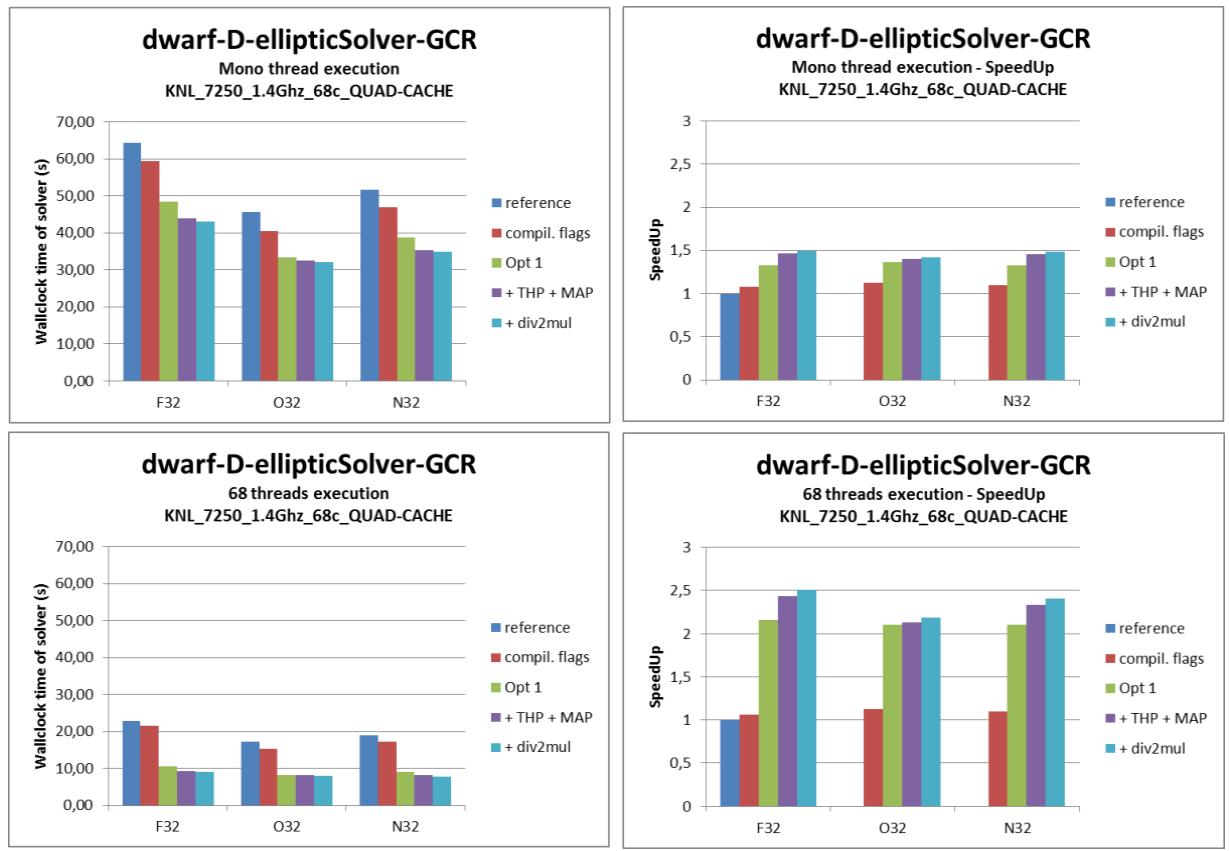

*Figure 33 - Elliptic solver GCR dwarf – Optimizations results on both mono and multi-threads executions (Xeon Phi)*

### 9.1.4  Summary

In this part, a brief profiling and benchmarking of the Elliptic solver GCR dwarf has been presented in addition with simple code optimization on the most time consuming function and their corresponding results. This original version of the code suffers from a low OpenMP efficiency and may be an issue concerning the MPI parallelisation. The performed optimizations led to a speedup of 2.5x on a multi-thread execution. As a conclusion, improving the OpenMP efficiency and data locality represents the more relevant optimization tracks to achieve better performance. This will imply more intrusive optimization.

## Document History

| Version | Author(s) | Date | Changes |
|---------|-----------|------|---------|
| 0.1 | Peter Messmer (NVIDIA) | 2017/11/26 | Initial version |
| 0.5 | Peter Messmer (NVIDIA) | 2017/12/12 | Major additions; Submitted for review |
| 1.0 | Peter Messmer (NVIDIA | 2017/12/22 | Final Version |

## Internal Review History

| Internal Reviewers | Date | Comments |
|--------------------|------|----------|
| Mike Gillard (LU) | 18/12/2017 | Approved with comments |
| Erwan Raffin (Bull) | 18/12/2017 | Approved with comments |
|  |  |  |
|  |  |  |

## Effort Contributions per Partner

| Partner | Efforts |
|---------|---------|
| NVIDIA | 8.1 |
| Bull/ATOS | 11.2 |
| Optalysys | 13.7 |
| **Total** | **33** |

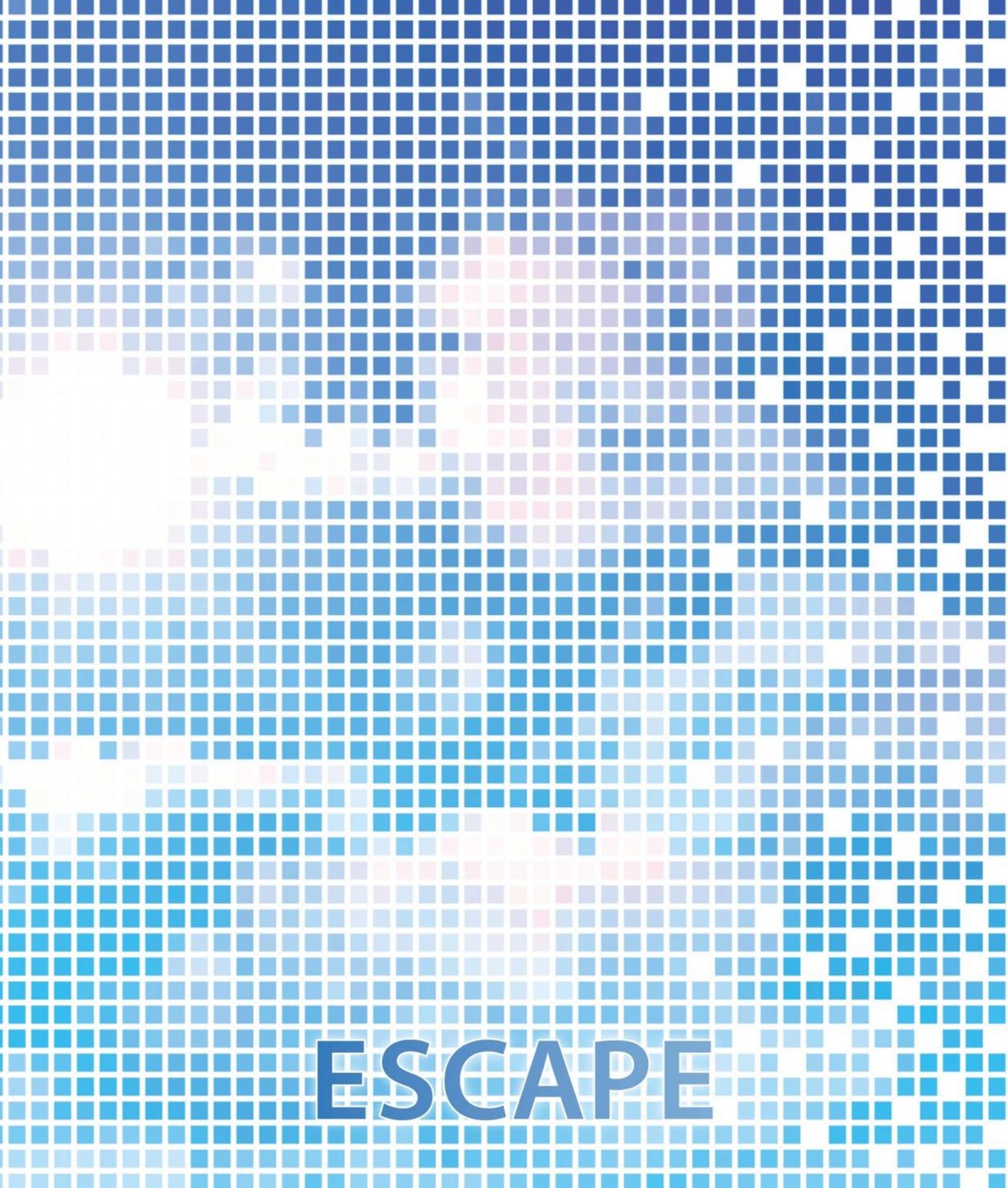